\documentclass[useAMS,usenatbib]{mn2e}
\usepackage{times}
\usepackage{amsmath}
\usepackage{graphicx}
\usepackage{color}
\newcommand{\mpc}{\mbox{$ h^{-1} \rmn{Mpc}$}}
\newcommand{\lya}{\mbox{$\rmn{Ly}\alpha$}}
\newcommand{\ha}{\mbox{$\rmn{H}\alpha$}}
\newcommand{\galform}{\texttt{GALFORM}}
\newcommand{\euclid}{\texttt{Euclid}}

\newcommand{\bau}{\texttt{Bau05}\,}
\newcommand{\bow}{\texttt{Bow06}\,}
\newcommand{\baur}{\texttt{Bau05(r)}}
\newcommand{\bowr}{\texttt{Bow06(r)}}
\newcommand{\bowd}{\texttt{Bow06(d)}}

\newcommand{\funits}{\mbox{$[\rmn{erg} \ \rmn{s}^{-1} \ \rmn{cm}^{-2}]$}}		
\newcommand{\ewobs}{\mbox{$EW_{\rmn{obs}}$}}						

\newcommand{\om}{\mbox{$\Omega_{\rm{m}}$}}
\newcommand{\ol}{\mbox{$\Omega_{\rm{\Lambda}}$}}

\newcommand{\veff}{\mbox{$V_{\rm eff}$}}

\newcommand{\logfha}{\mbox{${\rm log}(F_{H\alpha}\funits)$}}
\newcommand{\loglha}{\mbox{${\rm log}(L_{H\alpha}\funits)$}}

\newcommand{\xid}{\mbox{$\xi(r_{\sigma},r_{\pi})$}}
\newcommand{\dndz}{\mbox{${\rm d}N/{\rm d}z$}}
\title[\ha\ versus H-band selection in future redshift surveys]
{Probing dark energy with future redshift surveys: A comparison of 
emission line and broad band selection in the near infrared} 
\author [A. Orsi et al.]
{Alvaro Orsi$^1$\thanks{Email: alvaro.orsi@durham.ac.uk}, C. M. Baugh$^1$, C. G. Lacey$^1$, A. Cimatti$^2$, Y. Wang$^3$, G. Zamorani$^4$\\
1. Institute for Computational Cosmology, Department of Physics, University of Durham, South Road, Durham DH1 3LE, UK. \\
2. Dipartimento di Astronomia, Universita di Bologna, via Ranzani 1, I-40127, Bologna, Italy.\\
3. Homer L. Dodge Department of Physics \& Astronomy, University of Oklahoma, 440 W. Brooks St., Norman, OK 73019, USA.\\
4. INAF, Osservatorio Astronomico di Bologna, via Ranzani 1, I-40127 Bologna, Italy.\\}
\begin{document}
\maketitle
\begin{abstract}
Future galaxy surveys will map the galaxy distribution in the redshift 
interval $0.5<z<2$ using near-infrared cameras and spectrographs. The 
primary science goal of such surveys is to constrain the nature of the 
dark energy by measuring the large-scale structure of the Universe. 
This requires a tracer of the underlying dark matter 
which maximizes the useful volume of the survey. We investigate 
two potential survey selection methods: 
an emission line sample based on the \ha\ line and a sample selected 
in the H-band. We present predictions for the abundance and clustering 
of such galaxies, using two published versions of the \galform\ galaxy 
formation model. Our models predict that \ha\ selected galaxies tend 
to avoid massive dark matter haloes and instead trace the surrounding 
filamentary structure; H-band selected galaxies, on the other hand, 
are found in the highest mass haloes. This has implications for the 
measurement of the rate at which fluctuations grow due to gravitational 
instability. We use mock catalogues to compare the effective volumes 
sampled by a range of survey configurations. 
To give just two examples: a redshift survey down to $H_{\rm AB}=22$ samples 
an effective volume that is $\sim 5-10$ times larger than that probed by an 
\ha\ survey with $\logfha > -15.4$;  a flux limit of at least $\logfha = -16$ 
is required for an \ha\ sample to become competitive in effective volume.  
\end{abstract}


\section{Introduction}

A number of approaches have been proposed to uncover the nature of the 
accelerating expansion of the Universe which involve measuring the large 
scale distribution of galaxies \citep[e.g ][]{albrecht06,peacock06}. 
The ability of galaxy surveys to discriminate between competing  
models depends on their volume. 
Once the solid angle of a survey has been set, the useful volume 
can be maximised by choosing a tracer of the large-scale structure 
of the Universe which can effectively probe the geometrical volume. 
This depends on how the abundance of tracers drops with increasing 
redshift, and how much of this decline is offset by an increase in 
the clustering amplitude of the objects. 

Several wide-angle surveys have probed the redshift 
interval between $0<z<1$ \citep[e.g ][]{colless03,york00,cannon06,blake09}. 
The next major step up in volume will be made when the range from 
$0.5 <z <2$ is opened up with large near-infrared cameras and 
spectrographs which are mounted on telescopes able to map solid angles 
running into thousands of square degrees. From the ground, this part 
of the electromagnetic spectrum is heavily absorbed by water vapour in 
the Earth's atmosphere and affected by the strong atmospheric
OH emission lines. A space mission to construct an all-sky map 
of galaxies in the redshift range $0.5<z<2$ would have a significant 
advantage over a ground based survey in that the sky background 
in the near-infrared (NIR) is around 500 times weaker in space than 
it is on the ground. 

An important issue yet to be resolved for a galaxy survey 
extending to  $z \sim 2$ is the construction of the sample and the 
method by which the redshifts will be measured. One option is to use 
slitless spectroscopy and target the \ha\ emission line.  \ha\ is 
located at a restframe wavelength of $\lambda = 6563$\AA{}, which, for galaxies at $z>0.5$, 
falls into the near-infrared part of the electromagnetic spectrum 
\citep{thompson96,mccarthy99,hopkins00,shim09}. \ha\ emission is powered 
by UV ionizing photons from massive young stars. The only source of 
attenuation is dust, which is less important at the wavelength 
of \ha\ than it is for shorter wavelength lines. This makes \ha\ a more 
direct tracer of galaxies which are actively forming stars than 
other lines such as \lya, OII, OIII, H$\beta$ or H$\gamma$, which 
suffer from one or more sources of attenuation (i.e. dust, stellar 
absorption, resonant scattering) and which are more sensitive to 
the metallicity and ionisation state of the gas. The second option 
is to use some form of multi-slit spectrograph to carry out a redshift 
survey of a magnitude limited sample. The use of a slit means that 
unwanted background is reduced, allowing fainter galaxies to be targetted. 
Also, it is easier to identify which spectrum belongs to which galaxy 
with a slit than it is with slitless spectroscopy. 
Targets could be selected in the H-band at an effective wavelength 
of just over  1 micron, which is around the centre of the near infrared
wavelength part of the spectrum.

Space missions designed to carry out redshift surveys like the ones outlined 
above are currently being planned and assessed on both sides of the Atlantic. 
At the time of writing, the European Space Agency is conducting a Phase A 
study of a mission proposal called Euclid \footnote[1]{http://sci.esa.int/science-e/www/object/index.cfm?fobjectid=43226},
one component of which is a 
galaxy redshift survey. Both of the selection techniques mentioned above are being 
evaluated as possible spectroscopic solutions. The slit solution for 
Euclid is based on a novel application of digital 
micromirror devices (DMDs) to both image 
the galaxies to build a parent catalogue in the H-band and to measure 
their redshifts (see Cimatti et~al. 2009 for further details about the 
Euclid redshift survey). A \ha\ mission is also being discussed in the USA
\footnote[2]{http://jdem.gsfc.nasa.gov/}. 
At this stage, the sensitivity of these missions is uncertain and subject 
to change. For this reason we consider a range of \ha\ flux limits 
and H-band magnitudes when assessing the performance of the surveys. 
The specifications and performance currently being 
discussed for these missions have motivated the range of fluxes 
that we consider. 

A simple first impression of the relative merits of different  
selections methods can be gained by calculating the effective volume 
of the resulting survey. This requires knowledge of the survey geometry 
and redshift coverage, along with the redshift evolution of the 
number density of sources and their clustering strength. 
In this paper we use published galaxy formation models to 
predict the abundance and clustering of different samples of 
galaxies in order to compute the effective volumes of  a range 
of \ha\ and H-band surveys. Observationally, 
relatively little is known about the galaxy population selected by 
\ha\ emission or H-band magnitude at $0.5 < z < 2$. 
Empirically it is possible to estimate the number density of sources 
from the available luminosity function data 
and, on adopting a suitable model, to use the limited clustering 
measurements currently available to infer the evolution of the number 
density and bias \citep{shioya08,morioka08,geach08}. 
Geach et~al. (2009), in a complementary study to this one, 
make an empirical estimate of the number density of \ha\ emitters, 
and combine this with the predictions of the clustering of these galaxies 
presented in this paper to estimate the efficiency with which 
\ha\ emitters can measure the large scale structure 
of the Universe. 

The outline of the paper is as follows: in Section~\ref{sec.models} we 
give a brief overview of the models. Some general properties of 
\ha\ emitters in the models, such as luminosity functions (LF), 
equivalent widths (EW) and clustering bias are presented in 
Section~\ref{sec.properties} as these have not been published 
elsewhere. In Section~\ref{sec.DE} we show how our models can be used 
to build mock survey catalogues. We analyse the differences in the 
clustering of \ha\ emitters and H-band selected galaxies and present 
an indication of the efficiency with which different surveys trace 
large-scale structure (LSS). Finally, we give our conclusions 
in Section~\ref{sec.conclusions}.

\section{The Models}
\label{sec.models}
In this paper we present predictions for the clustering of galaxy 
samples selected in the near-infrared using two published versions 
of the semi-analytic model \galform. An overview of the semi-analytical 
approach to modelling galaxy formation can be found in Baugh (2006). 
The \galform\ code is described in \citet{cole00} and \citet{benson03}. 
The two models considered in this paper are explained fully in the 
original papers, \citet{baugh05} (hereafter the \bau\ model) and 
\citet{bower06} (hereafter the \bow  model). A thorough description 
of the ingredients of the \bau\ model can also be found in 
Lacey et~al. (2008); detailed comparisons of the physical ingredients 
of the two models are given in Almeida et~al. (2007, 2008), 
\citet{gonzalez09} and \citet{gonzalez-perez09}. Here we give 
an overview of the main features of each model and refer the reader 
to the above references for further details. 

The models are used to calculate the properties of the galaxy 
population as a function of time, starting from the merger histories 
of dark matter haloes and invoking a set of rules and recipes to 
describe the baryonic physics. These prescriptions require parameter 
values to be set to define the model. These values are set 
by comparing the model predictions against observations 
of local galaxies. The \bau  and \bow models have many ingredients in 
common but differ in the way in which they suppress the formation 
of bright galaxies. Also, different emphasis was placed on reproducing 
various local datasets when setting the parameters of the two models. 
It is important to remember that our starting point here is the two ``off the 
shelf'' galaxy formation models, which were set up without reference 
to \ha\ or H-band observations. In view of this it is remarkable how close 
these models come to matching the observed \ha LFs and H-band counts and 
redshift distributions, as presented in the next sections.  

The \bau model uses a superwind to stifle the formation of bright 
galaxies. The rate of mass ejection is assumed to be proportional 
to the star formation rate. The superwind ejects baryons from 
small and intermediate mass 
haloes. The cooling rate in massive haloes is reduced because these 
haloes have a reduced baryon fraction, due to the operation of the 
superwind in their progenitors. The model assumes that star formation 
which takes place in bursts occurs with a top-heavy initial mass 
function (IMF). For each solar mass of stars formed, four times the number 
of Lyman continuum photons are produced in a starburst as would be 
made in a quiescent episode of star formation, in which stars are 
produced with a standard solar neighbourhood IMF \citep{kennicutt83a}. 
Highlights of the \bau\ model include matching the observed number 
counts and redshift distribution of sub-millimetre sources and the 
luminosity function of Lyman-break galaxies. The \bau\ model also 
successfully reproduces the abundance and properties (including 
clustering) of \lya\ emitters \citep{dell1,dell2,orsi08}.
 
The \bow  model, on the other hand, uses feedback from active 
galactic nuclei (AGN) to stop the formation of bright galaxies. 
The accretion of ``cooling flow'' gas directly onto a central supermassive 
black hole releases jets of energy which heat the hot gas, and greatly 
reduces the cooling flow (see Croton et~al. 2005). 
Hence the supply of cooling gas for star formation is switched off. 
The \bow\ model gives a good match to the bimodal nature of the 
colour distribution of local galaxies (Gonzalez et~al. 2009), 
to the abundance of red galaxies (Almeida et~al. 2007; 
Gonzalez-Perez et~al. 2009) and to the evolution of the stellar 
mass function (Bower et~al. 2006). 

\begin{figure*}
\centering
\includegraphics[width=15cm,angle=90]{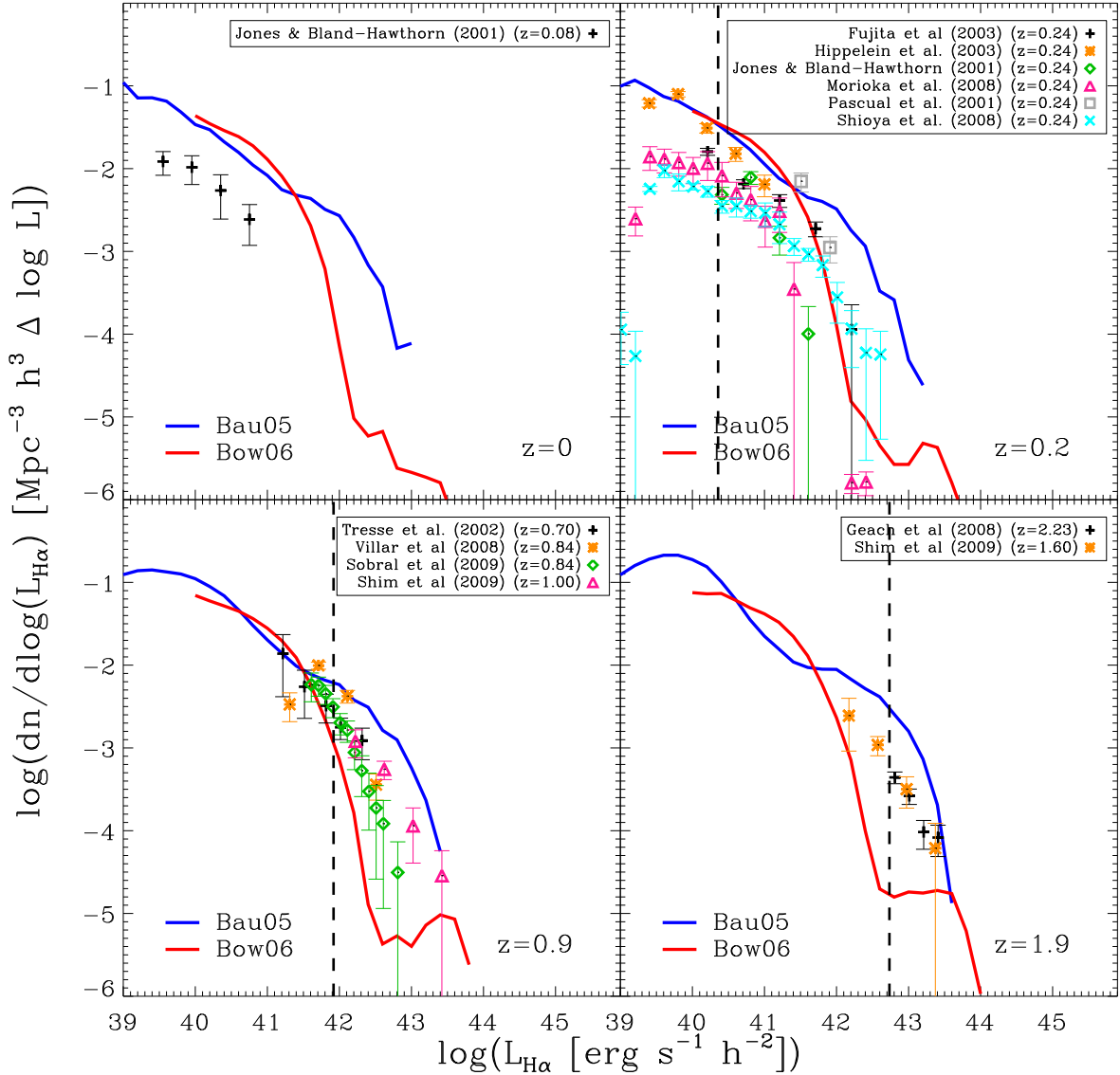}
\caption{
The \ha\ luminosity function, including attenuation by dust, 
at different redshifts. The blue curves show the predictions of 
the \bau\ model, whereas red curves show the \bow\ model. The 
observational estimates are represented by the symbols (see text 
for details). The redshift displayed in the bottom-right corner 
of each panel gives the redshift at which the \galform\ models 
were run. The vertical black dashed line shows the \ha\ luminosity 
corresponding to the flux $\logfha = -15.4$ for $z>0$, displayed
to show the expected luminosity limit of current planned space missions.
}
\label{fig.LF}
\end{figure*}

Other differences between the two models include: i) starbursts 
triggered by dynamically unstable disks in 
the \bow\ model; ii) a universal solar neighbourhood IMF in the \bow\ 
model; iii) the use of dark matter halo merger histories 
extracted from an N-body simulation in the \bow  model, whereas the 
\bau model uses Monte-Carlo generated trees; 
iv) a slightly different set of cosmological parameters ($\om = 0.3$, 
$\ol = 0.7$, $\Omega_b = 0.04$, $h = 0.7$, $\sigma_8 = 0.9$ for the \bau 
model, 
and $\om = 0.25$, $\ol = 0.75$, $\Omega_b=0.045$, $h = 0.73$, 
$\sigma_8 = 0.93$ for the \bow model).

The calculation of H-band flux and \ha\ line emission is the same in both models. 
The model predicts the star formation history of each galaxy, recording 
the star formation rate and the metallicity with which stars are made  
in each of the galaxy's progenitors. This allows a composite stellar population 
and spectral energy distribution to be built up. The model predicts the 
scale size of the galaxy and, through a chemical evolution model, the 
metal content of the disk and bulge. The H-band magnitude is computed 
by convolving the model galaxy spectral energy distribution with 
an H-band filter, appropriately shifted in wavelength if the galaxy is observed  
at $z>0$. The effect of dust extinction is taken into account by assuming 
that the dust and disk stars are mixed together (Cole et~al. 2000). The spectral 
energy distribution also gives the rate of production of Lyman continuum 
photons. Then, all of the ionizing photons are assumed to be absorbed 
by the neutral gas in the galaxy, and, by adopting case B recombination 
(Osterbrock 1989), the emissivity of the \ha\ line (and other emission 
lines) is computed. Here we assume that the attenuation of the \ha\ 
emission is the same as that experienced by the continuum at the 
wavelength of \ha\ . To predict the equivalent width (EW) of the \ha\ 
emission, we simply divide the luminosity of the line by the luminosity 
of the continuum around the \ha\ line.

\section{Properties of \ha\ emitters}
\label{sec.properties}
We first concentrate on the nature of \ha\ emitters in the models, 
which have not been discussed elsewhere for \galform\ , before 
examining the clustering of \ha\ and H-band selected samples 
in more detail in the next section.  In this section we present the 
basic predictions for the abundance, equivalent width distributions 
and clustering of \ha\ emitters.  Note that all the results presented 
here include the attenuation of the \ha\ emission by dust in the ISM 
at the same level experienced by the continuum at the wavelength 
of \ha\ . 

\subsection{The \ha\ luminosity function}
A basic prediction of the models is the evolution of the \ha\ 
luminosity function (LF). Fig.~\ref{fig.LF} shows the \ha\ LFs predicted 
by the two versions of \galform\ compared with observational data, 
over the redshift interval $0<z<2$. At each redshift plotted, 
the \bau\ model predicts a higher number density of \ha\ emitters 
than the \bow\ model for luminosities brighter than $\loglha \simeq 42$. 
This reflects two processes: the relative efficiency of the feedback 
mechanisms used in the two models to suppress the formation of bright 
galaxies, and the top-heavy IMF adopted in starbursts in the \bau\ model, 
which, for a galaxy with a given star formation rate, boosts the 
\ha\ flux emitted. The bright end of the \ha\ LF is dominated by 
bursting galaxies.

At faint luminosities, Fig.~\ref{fig.LF} shows that the predicted model 
LFs are more similar. At these luminosities, the star formation in both 
models predominantly takes place in galactic disks and produces stars 
with a standard IMF. For luminosities fainter than $\loglha \simeq 40$, 
the \bow\ model suffers from the limited mass resolution of Millennium Simulation 
halo merger trees (Springel et~al. 2005) compared with that of the Monte Carlo trees 
used in the \bau\ model \citep{helly03}.

The observational data shown in Fig.~\ref{fig.LF} comes from 
\citet{jones01} for $z\sim 0$; \citet{fujita03},\citet{hippelein03} ,\citet{jones01},
\citet{morioka08},\citet{pascual01},\citet{shioya08} for $z \sim 0.2$; \citet{tresse02},
\citet{villar08},\citet{sobral09},\citet{shim09} for $z \sim 0.9$ and \citet{geach08},\citet{shim09} 
for $z=2.2$. Most of this observational data has not been corrected 
by the authors for dust extinction, and hence it can be directly 
compared to the \galform\ predictions, which include dust attenuation. 
However, in some cases the data were originally presented after correction for an assumed
constant attenuation. In such cases we have undone this ``correction''. 
Hence, our comparison concerns the actual observed number of \ha\ emitters, 
which is the relevant quantity for assessing the performance of a redshift survey. 

In general both models overpredict the number of low luminosity 
\ha\ emitters at $z \leq 0.3$, as shown by Fig.~\ref{fig.LF}. 
At $z=0$, (upper-left panel in Fig.~\ref{fig.LF}), the amplitude 
of the LF in both models is larger, by almost an order of magnitude, 
than the \citet{jones01} data. A similar conclusion is reached at $z=0.2$ 
(upper-right panel in Fig.~\ref{fig.LF}), on comparing the models to most 
of the observational data. However, there is a significant scatter in observations of the 
faint end of the LF. At redshifts $z\ga 1$ (bottom panels in 
Fig.~\ref{fig.LF}), the models bracket the observational estimates, with 
the \bow\ model tending to underpredict the observational LF, whereas 
the \bau\ model over predicts it. Despite the imperfect agreement, 
these model LFs ``bracket'' the observed LFs for the redshifts relevant
to space mission surveys propsed, so we proceed to use them for the purposes
of this paper.

\begin{figure}
\centering
\includegraphics[width=13cm,angle=90]{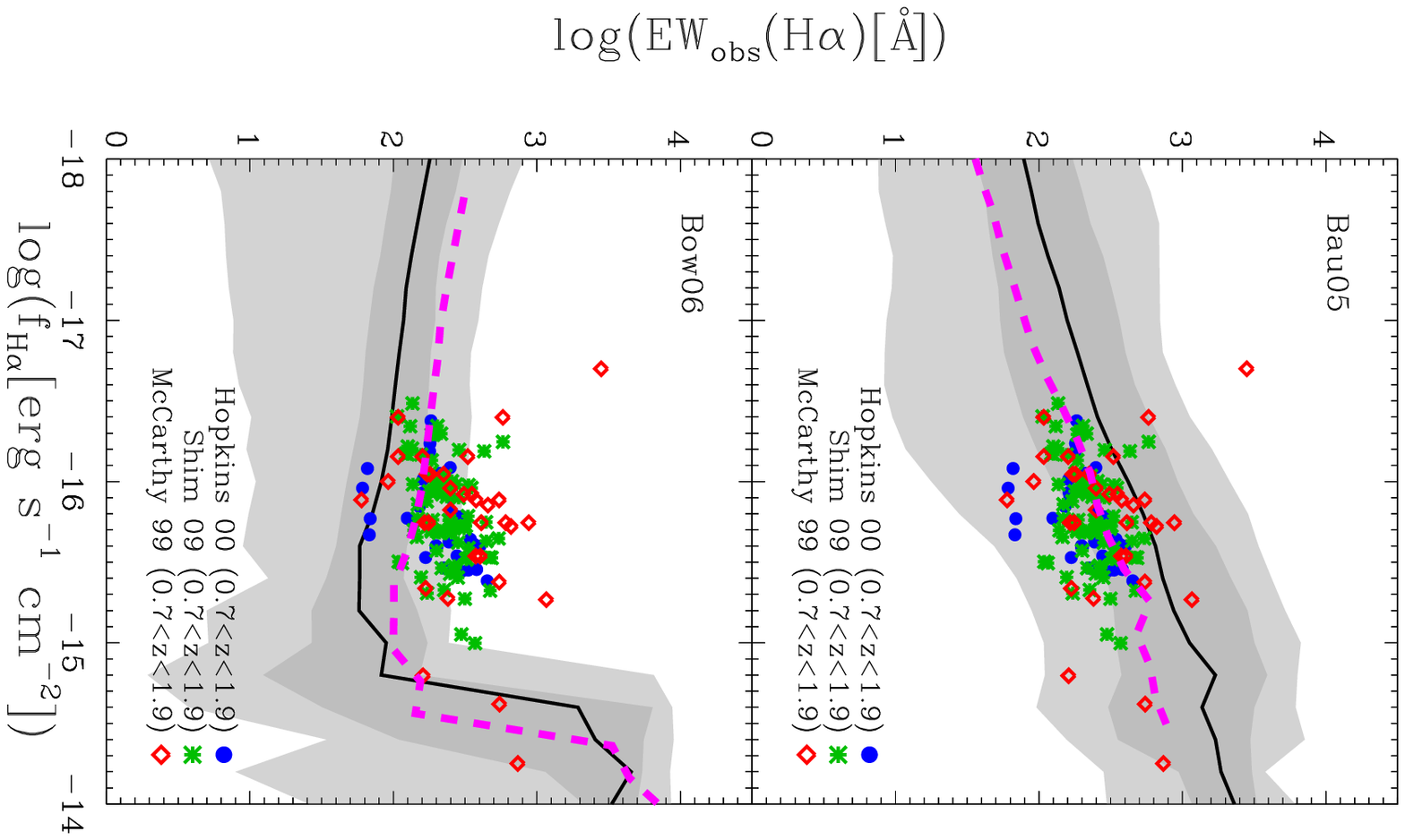}
\caption{
The distribution of \ha\ equivalent width in the observer frame 
as a function of \ha\ flux, over the redshift interval $0.7<z<1.9$.
The top panel shows the predictions of the \bau\ model  
and the bottom panel shows the \bow\ model, calculated as described 
in the text. The black line shows the median EW at each flux. The shaded 
regions enclose $68\%$ (dark grey) and $95\%$ 
(light grey) respectively of the \galform\ predictions around the median 
(black circles). The blue circles show observational data from 
\citet{hopkins00}, green asterisks show data from \citet{shim09} 
and red diamonds show data from \citet{mccarthy99}, as indicated 
by the key. The magenta dashed lines show the \galform\ predictions for
the median equivalent width after applying the empirically derived continuum 
flux and line luminosity rescalings described in Section~\ref{sec.DE}. 
}
\label{fig.ew} 
\end{figure}
\subsection{\ha\ equivalent width (EW) distribution}

Broadly speaking the EW of the \ha\ line depends on the current SFR in a galaxy  
(which determines the \ha\ emission), and its stellar mass (to which 
the continuum luminosity is more closely related). We compare the model predictions 
for the EW of \ha\ versus \ha\ flux with observational results 
in Fig.~\ref{fig.ew}. The observational data cover a wide redshift 
interval, $0.7<z<1.9$ \citep{mccarthy99,hopkins00,shim09}. 
In order to mimic the observational selection when generating model 
predictions, we go through the following two steps. First, we run 
the models for a set of redshifts covering the above redshift range. 
Second, we weight the \ewobs\ distribution at a given flux by 
\dndz, the redshift distribution of \ha\ emitters 
over the redshift range, to take into account the change in 
the volume element between different redshifts 
(see Section \ref{sec.DE} for details of the calculation of 
\dndz).

Fig.~\ref{fig.ew} shows the \ewobs\ distribution predicted by the 
\bau\ model (top panel) and the \bow\ model (bottom panel). The 
models predict different trends of \ewobs\ with \ha\ flux. 
In the \bau\ model, the typical EW increases with \ha\ flux, with a 
median value close to $\ewobs \sim 100$\AA{} at $\logfha = -18$, 
reaching $\ewobs \sim 2000$\AA{} at $\logfha = -14$. In contrast, 
the \bow\ model predicts a slight decline of \ewobs\ with \ha\ flux 
until very bright fluxes are reached, with median $\ewobs \sim 100$\AA{} 
in the range $\logfha = [-18,-15]$. For $\logfha>-15$, the \bow\ 
model predicts a sharp increase of the median $\ewobs$ to $ \sim 3000$\AA{}. The 
$95\%$ interval of the \ewobs\ found in \galform\ galaxies (the light 
grey region in Fig.~\ref{fig.ew}) covers almost 2 orders of magnitude 
in both models, except in the plateau found in the brightest bin of 
the \bow\ model, where the distribution covers 3 orders of magnitude. 
The \bau\ model matches the observed distribution of equivalent widths 
the best, particularly after the rescaling of continuum and line 
luminosities discussed in the next section (after which the median 
EW versus \ha\ distribution shifts from the solid black to the dashed magenta line). 
It is interesting to note that the ``shifted'' relations (see \S 4) give a better 
match to the observations for both models (although the \bau\ model remains a better fit),
particularly as the shift was derived with reference to the H-band galaxy number counts 
(for the continuum) and to the $z\sim1$ \ha\ LF, rather than to the EW data.   

\begin{figure}
\centering
\includegraphics[width=8cm,angle=90]{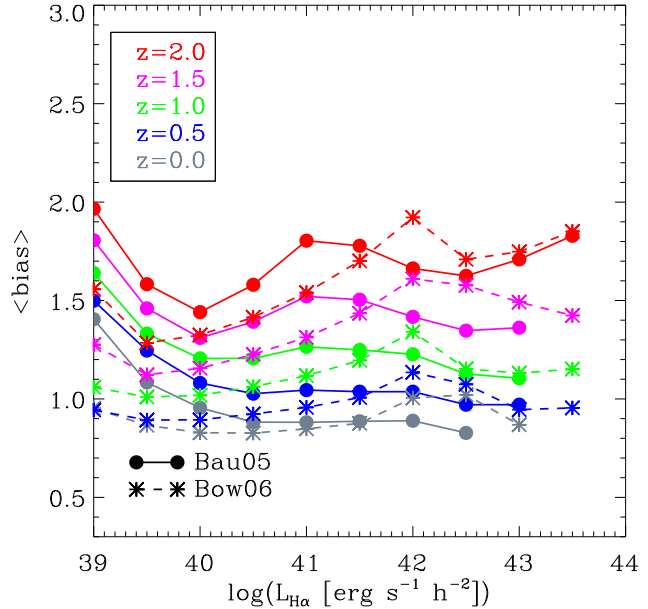}
\caption{The effective bias parameter as a function of \ha\ 
luminosity for redshifts spanning the range $0<z<2$. The \bau\ model 
results are shown using circles connected with solid lines and  
the \bow\ model results are shown with asterisks connected by 
dashed lines. Each colour corresponds to a different redshift, 
as indicated by the key. }
\label{fig.bias}
\end{figure}

\subsection{Clustering of \ha\ emitters: effective bias}

The clustering bias, $b$, is defined as the square root of the 
ratio of the galaxy correlation function to the correlation 
function of the dark matter \citep{kaiser84}. As we shall see in 
Section 4.3, the clustering bias is a direct input into the 
calculation of the effective volume of a galaxy survey, which 
quantifies how well the survey can measure the large scale structure 
of the Universe. Simulations show that the correlation functions 
of galaxies and dark matter reach an approximately constant 
ratio on large scales (see for example \citealt{angulo08a}; 
note, however, that small departures 
from a constant ratio are apparent even on scales in excess of 
$100 h^{-1}$Mpc). 

In this section we compute the effective bias of samples of 
\ha\ emitting galaxies. There are theoretical prescriptions 
for calculating the bias factor of dark matter haloes as a 
function of mass and redshift \citep{cole89,mo96,sheth01}. 
These have been extensively tested against the clustering of 
haloes measured in N-body simulations and have been found to be 
reasonably accurate \citep{gao05,wechsler06,angulo08b}. Here
we use \citet{sheth01}. 
The effective bias is computed by integrating over the halo mass 
the bias factor corresponding to the dark matter halo which hosts a galaxy 
multiplied by the abundance of the galaxies of the chosen luminosity
\citep[see, for example][]{baugh99,dell2,orsi08}. 
\begin{table}
\centering
\begin{tabular}{@{}lccc}
\hline
Model & $C_{H\alpha}$ & $C_{\rm cont}$ \\
\hline
Bau05 & 0.35 & 0.73 \\
Bow06 & 1.73 & 0.42 \\
\hline
\hline
\end{tabular}
\caption{Luminosity rescaling factors for the \ha\ line and the 
stellar continuum. Column 2 shows $C_{H \alpha}$, the factor used 
to adjust the predicted \ha\ flux as described in the text. This factor is 
only applied to the \ha\ line. Column 3 shows $C_{cont}$, the correction 
factor applied to the stellar continuum, as derived by forcing the model 
to match the observed H-band counts at $H_{\rm AB}=22$. This factor is 
applied to the entire stellar continuum of the model galaxies.} 
\label{table.offset}
\end{table}

Fig.~\ref{fig.bias} shows the predicted galaxy bias, $b_{\rm eff}$, 
as a function of \ha\ luminosity over the redshift interval $0<z<2$.
There is a clear increase in  the value of the effective 
bias with redshift; at $\loglha=40$, 
$b_{\rm eff} \approx 0.8$ at $z=0$, compared with $b_{\rm eff} \approx 1.5$ 
at $z=2$. Both models show an upturn in the effective bias with 
decreasing luminosity faintwards 
of $\loglha=40$. There is little dependence of bias on luminosity 
brightwards of $\loglha=40$, up to $z=2$. The predictions of the two 
models for the effective bias are quite similar. 
There are currently few observational measurements of the clustering of \ha\ 
emitters. \citet{geach08} inferred a spatial correlation 
length of $r_{0} = 4.2^{+0.4}_{-0.2}h^{-1}$Mpc for their sample of 
55 \ha\ emitters at $z=2.23$. This corresponds to a bias 
of $b\approx 1.7$ in the \bau\ model cosmology, which is in very good 
agreement with the predictions plotted in Fig.~\ref{fig.bias}.

\section{The effectiveness of redshift surveys for measuring dark energy}
\label{sec.DE}

In this section we assess the relative merits of using \ha\ or H-band 
selection to construct future redshift surveys aimed at measuring the dark energy
equation of state. The first step is to produce a mock catalogue 
that can reproduce currently available observations. We discuss 
how we do this in Section~4.1. We then present predictions for the 
clustering of \ha\ emitters and H-band selected galaxies in Section~4.2. 
We quantify the performance of the two selection methods in terms of 
how well the resulting surveys can measure the large-scale structure of 
the Universe in Section 4.3.

\subsection{Building accurate mock catalogues}
\label{sec.mocks}

\begin{figure}
\centering
\includegraphics[width=9cm,angle=90]{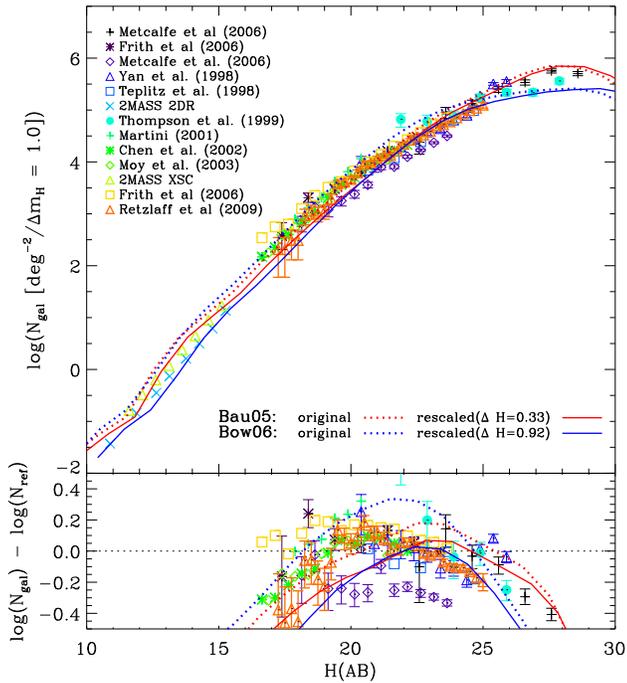}
\caption{
Number counts in the H band. The upper panel shows the differential counts 
on a log scale. The lower panel shows the counts after dividing by a power 
law $N_{\rm ref} \propto H_{\rm AB}^{0.32}$  to expand the dynamic range on the y-axis. The symbols show the 
observational data, as shown by the key in the upper panel. The lines show 
the model predictions. The dotted lines show the original \galform\ 
predictions for the \bau\ model (blue) and the \bow\ model (red). 
The solid curves show the rescaled \galform\ predictions after rescaling 
the model galaxy luminosities to match the observed number counts at 
$H_{\rm AB}=22$. 
}
\label{fig.n_m}
\end{figure}

\begin{figure}
\centering
\includegraphics[width=8cm,angle=90]{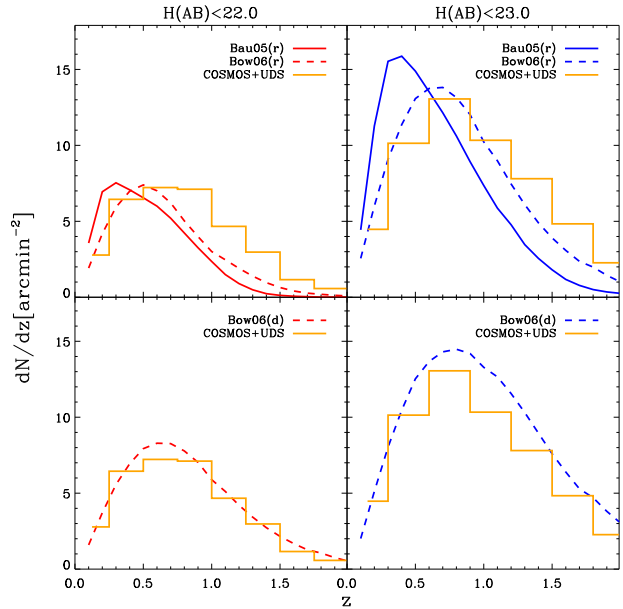}
\caption{
The redshift distribution of galaxies with $H_{\rm AB}=22$ (left
column) and $H_{\rm AB}<23$ (right column). 
The top panels show the predictions after rescaling the model 
luminosities to better match the number counts as explained in the text.
Red and blue lines show the model predictions for $H_{\rm AB}<22$ and $H_{\rm AB}<23$
respectively. Solid lines show the \baur\ model and the dashed lines show
the \bowr\ model. 
The lower panel shows the redshift distribution obtained from the \bow\ model 
by diluting the galaxies, randomly selecting 0.63 of the sample, the \bowd\ 
model (recall this is a purely illustrative case with no physical basis; 
see \S 4.1.1). In both panels, the histogram shows an estimate of the 
redshift distribution derived from spectroscopic observations in the 
COSMOS and UDF fields \citep[][; Euclid-NIS Science Team, 
private communication]{cirasuolo08}.
}
\label{fig.nzhband}
\end{figure}

Our goal in this section is to build mock catalogues for future 
redshift surveys which agree as closely as possible with currently 
available observational data. We have already seen that the models 
are in general agreement with observations of the \ha\ luminosity 
function, and will see in the next subsection how well the models match  
the H-band number counts. In our normal mode of operation, we set the 
model parameters with reference to a subset of local observations and 
see how well the model then agrees with other observables. This allows 
us to test the physics of the model; if the model cannot reproduce a  
dataset adequately, perhaps some ingredient is missing from the model 
\citep[e.g. for an application of this principle to galaxy clustering, 
see][]{kim09}. Here our primary aim is not to develop our 
understanding of galaxy formation physics but to produce a synthetic 
catalogue which resembles the real Universe as closely as possible. 
To achieve this end we allow ourselves the freedom to rescale the 
model stellar continuum and emission line luminosities, independently. 
This preserves the ranking of the model galaxies in luminosity. 
This approach is more powerful than an empirical model as we retain 
all of the additional information predicted by the semi-analytical model, 
such as the clustering strength of the galaxies.
Hereafter we will refer to the adjusted \bau\ and \bow\ models as 
\baur\ and \bowr\, respectively, to avoid confusion. We also consider 
a sparsely sampled version of the \bow\ model, which we refer to 
as \bowd\ (see \S 4.1.1). 

\subsubsection{H-band selected mock catalogues}

In Fig.~\ref{fig.n_m}, we first compare the model predictions without 
{\it any} rescaling of the luminosities against a compilation of 
observed number counts in the H-band, kindly provided by Nigel Metcalfe. 
Observational data are taken from the following sources, shown with different symbols: Black plus-signs from \citet{metcalfe06}; purple asterisks from \citet{frith06}; purple diamonds from \citet{metcalfe06}; blue triangles from \citet{yan98}; blue squares from \citet{teplitz98}; cyan crosses from the second data release of the 2MASS Survey \footnote[1]{http://www.ipac.caltech.edu/2mass/releases/second/\#skycover}; green circles from \citet{thompson99}; green plus-signs from \citet{martini01}; green asterisks from \citet{chen02}; green diamonds from \citet{moy03}; green triangles from the 2MASS extended source catalogue\footnote[2]{http://www.ipac.caltech.edu/2mass/releases/allsky/doc/sec2\_3d3.html}, orange squares from \citet{frith06}, and orange triangles from \citet{retzlaff09}\\

\begin{figure}
\centering
\includegraphics[width=7.5cm,angle=90]{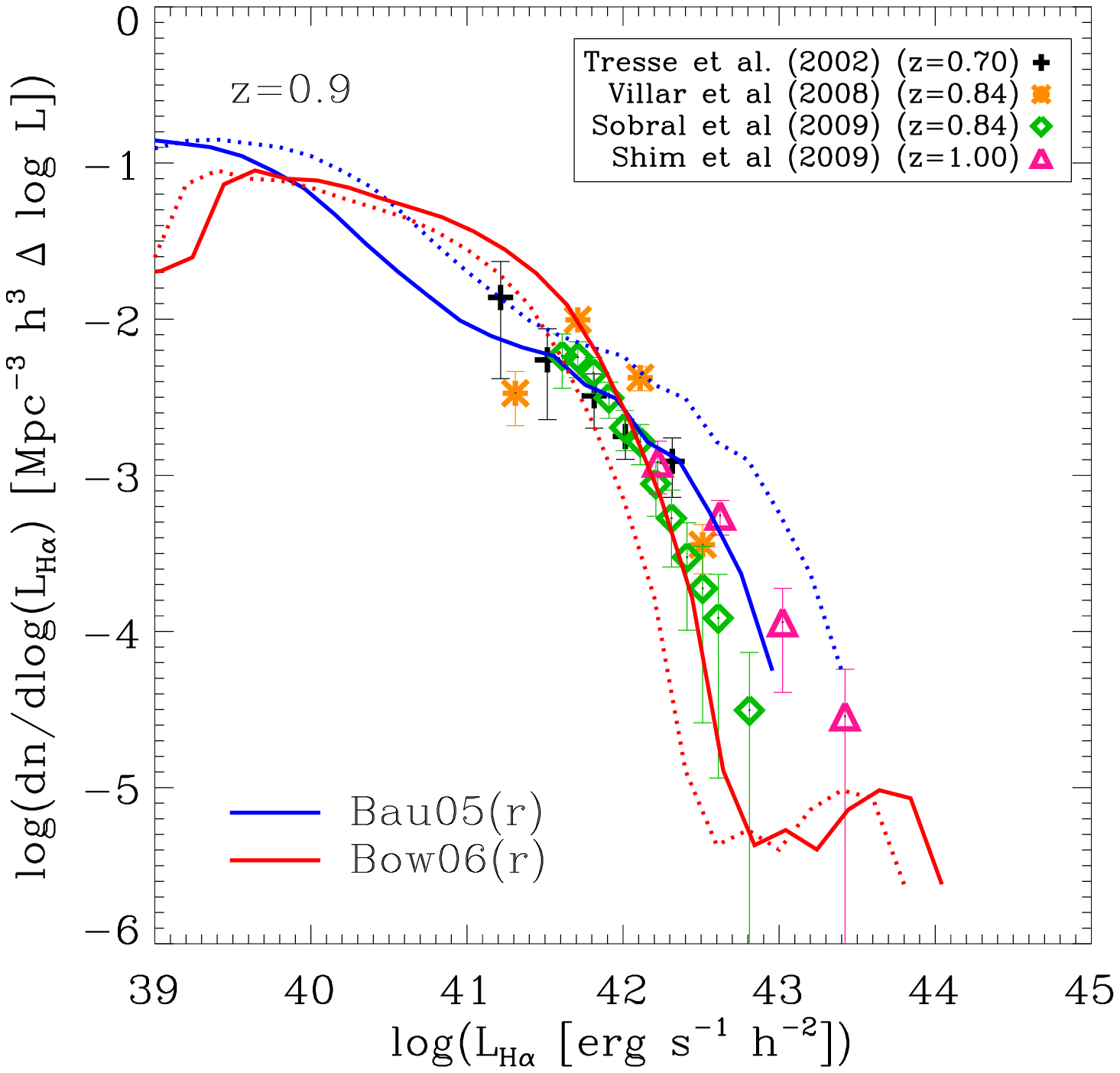}
\caption{
The \ha\ LF at $z=0.9$. The symbols show observational data, with the 
sources indicated in the key. The dotted curves show the original 
predictions for the \ha\ luminosity function, as plotted in Fig.~\ref{fig.LF}.
The solid curves show the model predictions after rescaling the \ha\ 
luminosity to better match the observed LF at $\loglha = 42$, which 
corresponds to a flux limit of $\logfha = -15.3$ at this redshift.  
}
\label{fig.LF_offset}
\end{figure}
\begin{figure}
\centering
\includegraphics[width=12cm,angle=90]{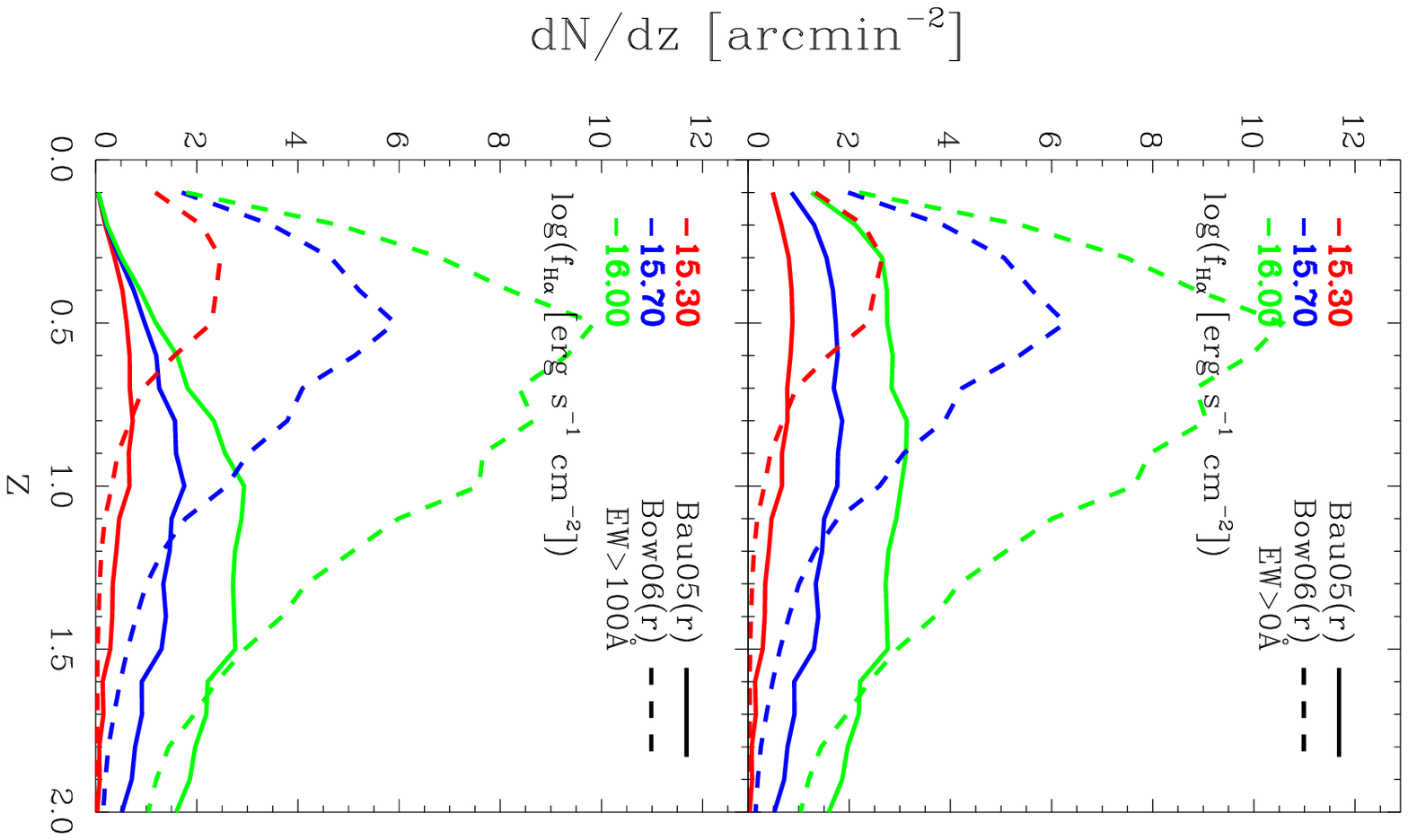}
\caption{
The redshift distribution of \ha\ selected galaxies for 3 different flux 
limits: $\logfha >$ -15.3, -15.7 and -16.0 shown in red, blue and green 
respectively. The solid lines show the \baur\ prediction and the dashed 
lines show the \bowr\ predictions. In the top panel, galaxies 
contributing to the redshift distribution have no cut imposed on the 
equivalent width of \ha\ . In the bottom panel, the model galaxies have to 
satisfy the \ha\ flux limit and a cut on the observed equivalent width 
of \ha\ of $\ewobs>100$\AA{}.}
\label{fig.nz}
\end{figure}

There is a factor of three spread in the observed counts around 
$H_{\rm AB}=20-22$. The unscaled models agree quite well with the observations 
at $H_{\rm AB}=20$ but overpredict the counts at $H_{\rm AB}=22$, the likely  
depth of a slit-based redshift survey from space. There are 
two ways in which the model predictions can be brought into better agreement 
with the observed counts at $H_{\rm AB}=22$; first, by rescaling the 
luminosities of the model galaxies to make them fainter in the H-band 
or second, by artificially reducing, at each magnitude, the number density 
of galaxies. 
The first correction could be explained as applying extra dust extinction 
to the model galaxies; as we will see later on, the typical redshift of the 
galaxies is $z \sim 0.5$--$1$, shifting the observer frame H into the 
rest frame R to V-band. 
The second correction has no physical basis and is equivalent to taking 
a sparse sampling of the catalogue at random, i.e. making a dilution of 
the catalogue. Galaxies are removed at random without regard to their 
size or redshift. 
(Note that the dissolution of galaxies invoked by Kim et~al. 2009 only 
applies to satellite galaxies within haloes, and is mass dependent, and 
hence is very different from the random dilution applied here.)
The motivation behind the second approach is that the shape of the original 
redshift distribution of the model is preserved. As we shall see, the 
first approach, rescaling the model galaxy luminosities, produces a 
significant change in the shape of the predicted redshift distribution.  

It is worth remarking in passing that the semi-analytical models used 
here have already been compared to the observed counts in the K-band 
\citep{gonzalez-perez09}. The \bow\ model was found to agree very well 
with the K-band observations whereas the \bau\ model underpredicted the 
counts by up to a factor of three. This is a somewhat different impression 
about the relative merits of the models from that reached on comparing 
to the observed H-band counts, which is surprising given the proximity of 
the bands and the similarity in the masses of the stars which 
dominate the light from the composite stellar populations at 
these wavelengths. 

The agreement with the observed counts is improved at $H_{\rm AB}=22$ 
by shifting the \bow\ galaxy magnitudes faintwards by $+0.92$ magnitudes; 
the \bau\ model requires a more modest dimming of $+0.33$ magnitudes (see 
Table ~\ref{table.offset}).

The redshift distribution of H-band selected galaxy samples provides a further 
test of the models. In Fig.~\ref{fig.nzhband}, the model predictions are 
compared against an estimate of the redshift distribution 
compiled using observations from the COSMOS survey and the Hubble 
Ultra-Deep Field for $H_{\rm AB}<22$ and $H_{\rm AB}<23$ 
(\citealt{cirasuolo08}; Cirasuolo, Le Fevre and McCracken, private 
communication). If we focus on the lower panels first, which 
shows \dndz\ in the randomly diluted \bow\ model, denoted as 
\bowd, it is apparent that the original \bow\ model predicted 
the correct shape for the redshift 
distribution of sources, but with simply too many galaxies at each 
redshift. In the upper panel of Fig.~\ref{fig.nzhband}, we see that 
the models with the shifted H-band luminosities give shallower 
redshift distributions than the observed one. The difference between 
the predicted \dndz\ after dimming the luminosities or diluting the 
number of objects has important implications for the number density 
of galaxies as a function of redshift, which in turn is important 
for the performance of a sample in measuring the large-scale structure 
of the Universe. 

\subsubsection{\ha-selected mock catalogues}

The original model predictions for the \ha\ luminosity function were presented 
in Fig.~\ref{fig.LF}. The models cross one another {\it and} match 
the observed \ha\ LF at a luminosity of $\loglha \sim 41.5 $. At $z=0.9$, 
this corresponds to a flux of $\logfha = -15.8$. The flux limit attainable 
by Euclid is likely to be somewhat brighter than this, although the precise 
number is still under discussion. For this reason, we chose to force the 
models to agree with the observed \ha\ LF at $\loglha = 42$ at $z=0.9$, 
which corresponds to a flux limit of $\logfha = -15.3$ (see Fig.~\ref{fig.LF_offset}). 
Before rescaling, the model LFs differ by a factor of three at $\loglha \sim 41.5$. 
In the rescaling, the \ha\ line luminosity is boosted in the \bow\ model 
and reduced in the case of the \bau\ model (see Table~\ref{table.offset} 
for the correction factors used in both cases). The latter could be 
explained as additional dust extinction applied to the emission line, 
compared with the extinction experienced by the stellar continuum.
The former correction, a boost to the \ha\ luminosity in the \bow\ model, 
is harder to explain. This would require a boost in the production of 
Lyman-continuum photons (e.g. as would result on invoking a top-heavy IMF 
in starbursts or an increase in the star formation rate). This would 
require a revision to the basic physical ingredients of the model and 
is beyond the scope of the current paper. 
\begin{figure*}
\centering
\includegraphics[width=7cm]{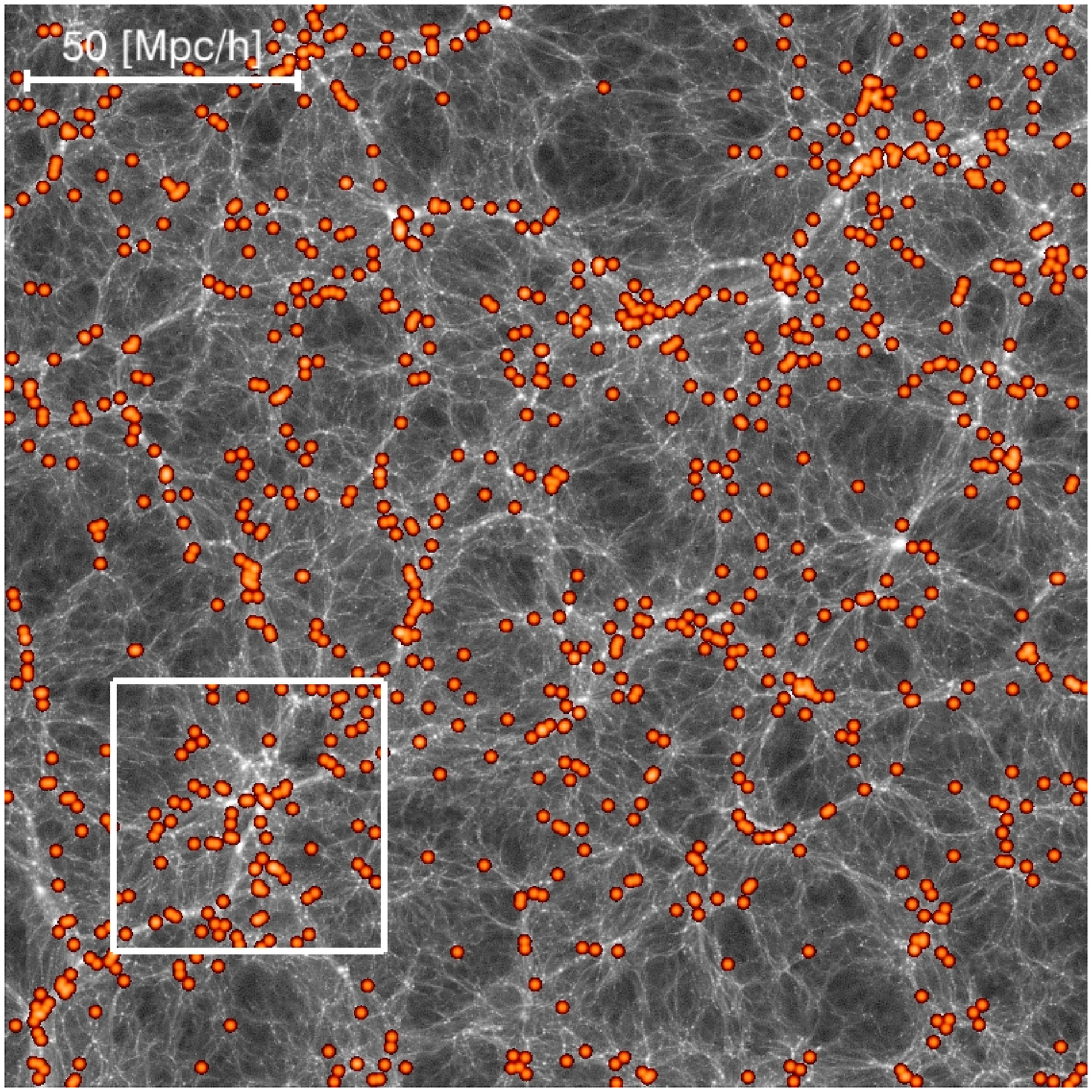}
\hspace{0.5cm}
\vspace{0.1cm}
\includegraphics[width=7cm]{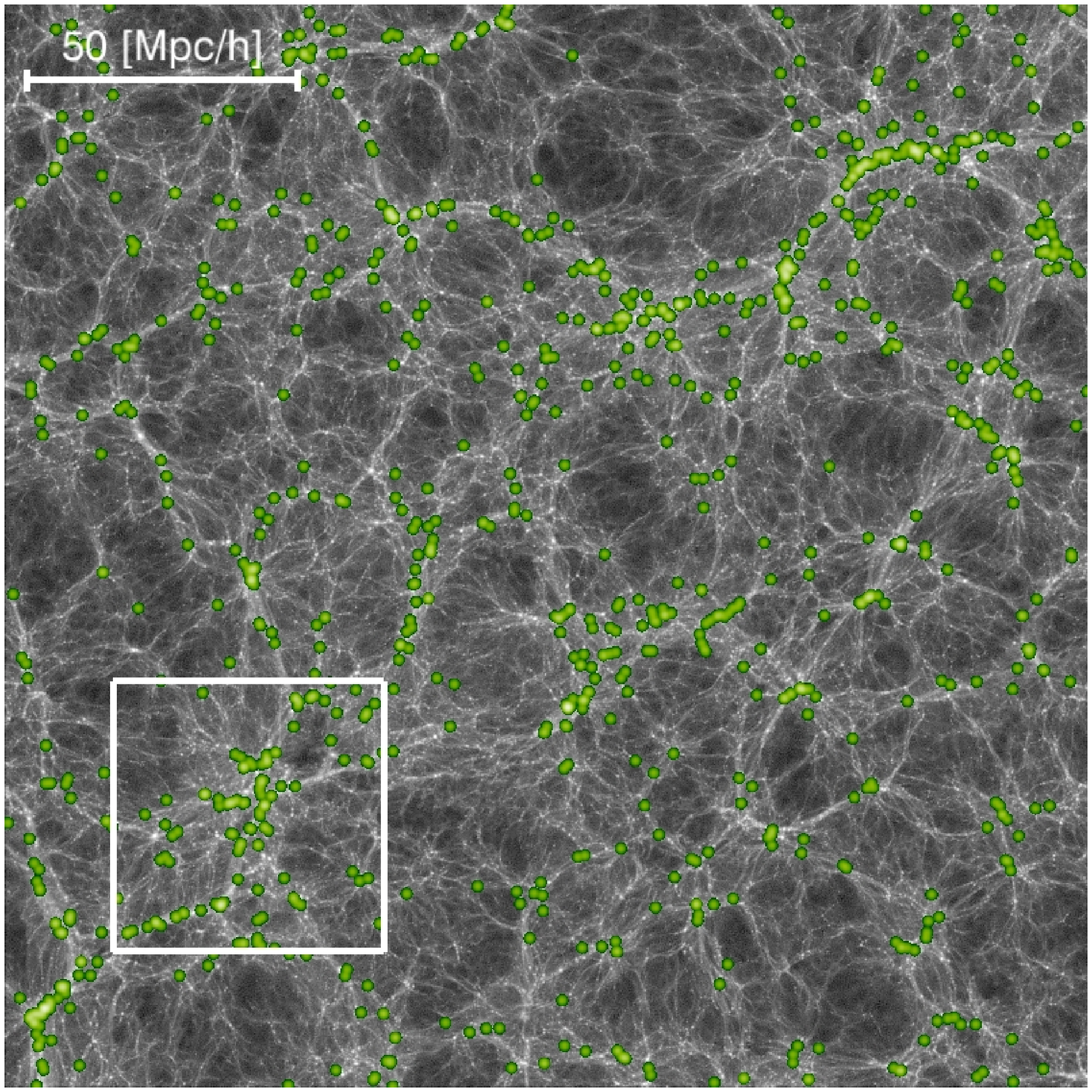}
\vspace{0.1cm}
\hspace{0.5cm}
\includegraphics[width=7cm]{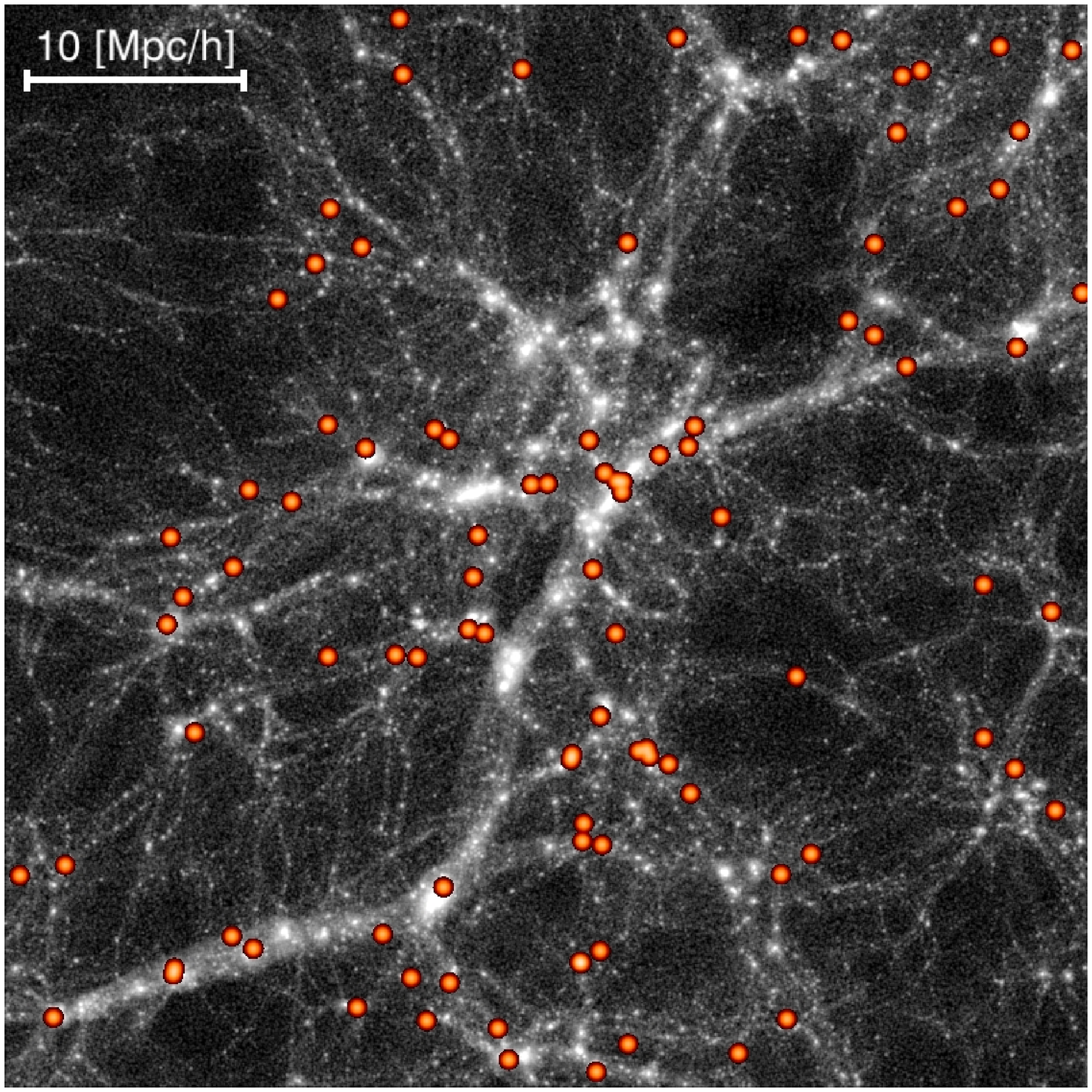}
\hspace{0.5cm}
\includegraphics[width=7cm]{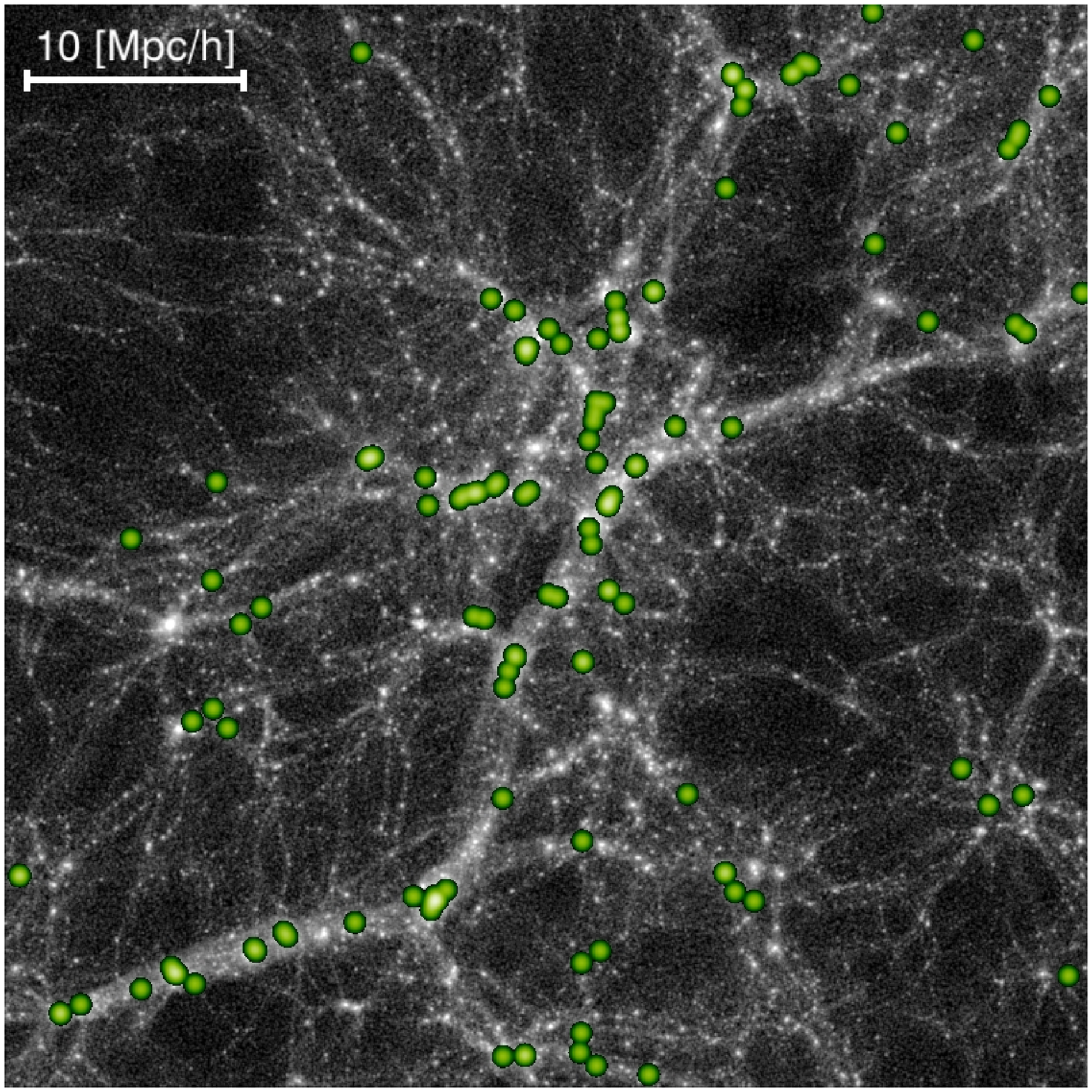}
\caption{
The spatial distribution of galaxies and dark matter in the \bowr\ model 
at $z=1$. Dark matter is shown in grey, with the densest regions shown 
with the brightest shading. Galaxies selected by their \ha\ emission 
with $\logfha>-16.00$ and and $\ewobs>100$\AA{} are shown in red in the 
left-hand panels. Galaxies brighter than $H_{\rm AB}=22$ are shown in green 
in the right-hand panels. Each row shows the same region from the 
Millennium simulation. The first row shows a slice of $200 \mpc$ on a 
side and $10\mpc$ deep. The second row shows a zoom into a region of 
$50\mpc$ on a side and $10\mpc$ deep, which corresponds to the white 
square drawn in the first row images. Note that all of the galaxies which 
pass the selection criteria are shown in these plots.}
\label{fig.dm}
\end{figure*}

After making this correction to the \ha\ line flux in the models, 
we next present the predictions for the redshift distribution of 
\ha\ emitters. Fig.~\ref{fig.nz} shows \dndz\ for flux 
limits of $\logfha=[-15.7,-16.0,-16.3]$. The redshift distribution 
of the \bowr\ model peaks around $z \sim 0.5$ and declines sharply 
approaching $z \sim 2$, whereas the \baur\ \dndz\  are much broader. 
The lower panel of Fig.~\ref{fig.nz} shows the redshift distribution 
after applying the flux limits and a cut on the observed equivalent 
width of $\ewobs=100$\AA{}. 
(Note that the \dndz\ is not sensitive to low EW cuts; similar results to the $\ewobs>0$ \AA{}
case are obtained with 10\AA{} in both models).
In the rescaled model, the equivalent width changes because the \ha\ line flux has been 
adjusted and because the continuum has been altered (by the same shift as 
applied to the H-band). Adding the selection on equivalent width results 
in a modest change to the predicted \dndz\ in the \bowr\ model. In the \baur\ 
model, the \dndz\ shifts to higher redshifts. 
There is no observational data on the redshift distribution of \ha\ emitters 
to compare against the model predictions. Geach et~al. (2009) make an 
empirical estimate of the redshift distribution, by fitting a model for 
the evolution of the luminosity function to observational data. 
The luminosity of the characteristic break in the luminosity function, 
$L_*$, is allowed to vary, while the faint end slope and normalisation 
are held fixed. The resulting empirical LF looks similar to the original 
\bau model at $z=0.9$, and the two have similar redshift distributions. 
The \ha\ redshift distributions in the \bowr\ models are shallower 
than the empirical estimate; the \baur\ model has a similar shape to the 
empirical redshift distribution, but with a lower normalisation.  It is 
important to realise that the approach of Geach et~al. is also model 
dependent, and the choices of model for the evolution of the luminosity 
function and of which observational datasets to match are not unique and 
will have an impact on the resulting form of the redshift distribution. 

\subsection{The clustering of \ha\ and H-band selected samples}

The semi-analytic galaxy formation model predicts the number of 
galaxies hosted by dark matter haloes of different mass. In the cases 
of \ha\ emission, which is primarily sensitive to ongoing star 
formation, and H-band light, which depends more on the number of 
long-lived stars, different physical processes determine the 
number of galaxies per halo. The model predicts contrasting  
spatial distributions for galaxies selected according to their \ha\ emission 
or H-band flux. We compare in Fig.~\ref{fig.dm} the spatial 
distribution of \ha\ emitters with fluxes $\logfha>-16$ and 
$\ewobs>100$\AA{} (red circles) 
with that of an H-band selected sample with $H_{\rm AB}<22$ 
(green circles), in the \bowr\ model which is set in the Millennium 
Simulation. The upper panels of Fig.~\ref{fig.dm} show how
the different galaxy samples trace the underlying cosmic web of dark matter.  
The lower panels of Fig.~\ref{fig.dm} show a zoom into a massive 
supercluster. There is a marked difference in how the galaxies trace 
the dark matter on these scales. The \ha\ emitters avoid 
the most massive dark matter structures. At the centre of massive 
haloes, the gas cooling rate is suppressed in the model due to AGN 
heating of the hot halo. This reduces the supply of gas for star 
formation and in turn cuts the rate of production of Lyman continuum 
photons, and hence the \ha\ emission. The H-band selected galaxies, 
on the other hand, sample the highest mass dark matter structures. 

To study the difference in the spatial distribution of galaxies 
in a quantitative way, we compare the clustering predictions from 
the models with observational data. Instead of computing the correlation 
function explicitly, we use the same method explained in Section 3.3 
to calculate the effective bias and use this to derive the correlation length, $r_0$, a measure of the 
clustering amplitude, which we define as the pair separation at which 
the correlation function equals unity.
The correlation function of galaxies, $\xi_{\rm gal}$, is related to 
the correlation function of dark matter, $\xi_{\rm dm}$, by 
$\xi_{\rm gal} = b^2\xi_{\rm dm}$. The effective bias is {\it approximately} 
constant on large scales (e.g. Angulo et~al. 2008a). We use the 
\citet{smith03} prescription to generate a nonlinear matter power 
spectrum in real space. This in turn is Fourier transformed to 
obtain the two-point correlation function of the dark matter, 
$\xi_{\rm dm}$. We can then derive $\xi_{\rm gal}$ for any survey 
configuration by multiplying $\xi_{\rm dm}$ by the square of the 
effective bias, and then we read off the correlation length as 
the scale at which the correlation function is equal to unity. 
\begin{figure}
\centering
\includegraphics[width=12cm,angle=90]{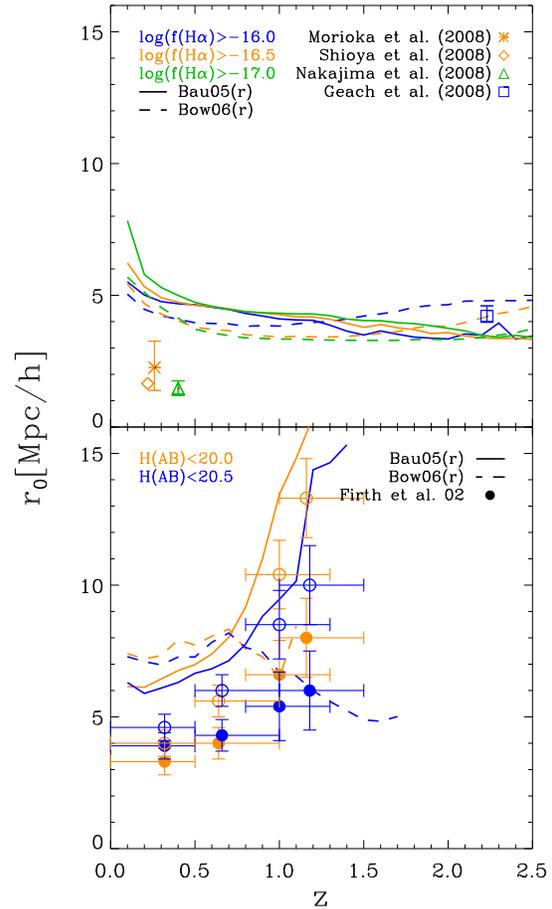}
\caption{
The correlation length, $r_0$, as a function of redshift 
for selected \ha\ and H-band samples. Solid and dashed lines 
show the predictions of the \bau\ and \bow\ models respectively. 
The top panel shows the predictions for different \ha\ limiting fluxes,  
$\logfha >[-16.0,-16.5,-17.0]$ in green, orange and blue respectively. 
Observational data is shown with symbols. The bottom panel shows the 
model predictions for $H_{AB}<[20.,20.5]$ in orange and blue respectively. 
In this case there are two sets of observational estimates, based on different 
assumptions for the evolution of clustering with redshift. 
}
\label{fig.r0}
\end{figure}

Fig.~\ref{fig.r0} shows the correlation length in comoving units 
for both \ha\ and H-band samples at different redshifts, compared 
to observational estimates. Differences in the bias predicted by the 
two models (as shown in 
Fig.~\ref{fig.bias}) translate into similar differences in $r_0$.  
The correlation length declines with increasing redshift for \ha\ emitters in the \baur\ model, 
since the increase of the effective bias with redshift is not strong 
enough to balance the decline of the amplitude of clustering of the dark 
matter. For the range of flux limits shown in the top panel of 
Fig.~\ref{fig.r0} ($-16 <\logfha<-17 $), $r_0$ changes from 
$\sim 5-7 \ \mpc$ at $z=0.1$ to $r_0\sim 3.5\ \mpc$ at $z=2.5$. 
On the other hand, the \bowr\ model shows a smooth increase of $r_0$ which
depends on flux and redshift. 
At bright flux limits $r_0$ evolves rapidly at high redshift, 
reaching $r_0 = 4.3 h^{-1}$Mpc at $z=2.5$. At fainter luminosities 
the change in correlation length with redshift is weaker.

The currently available observational estimates of the clustering 
of near infrared selected galaxy samples mainly come from angular 
clustering. A number of assumptions are required in order to derive 
a spatial correlation length from the angular correlation function. 
First, a form must be adopted for the distribution of sources in 
redshift. Second, some papers quote results in terms of proper 
separation whereas others report in comoving units. Lastly, an 
evolutionary form is sometimes assumed for the correlation function 
\citep{groth77}. In this case, the results obtained for the 
correlation length depend upon the choice of evolutionary model. 

Estimates of the correlation length of \ha\ emitters are available 
at a small number of redshifts from narrow band surveys, as shown in 
Fig.~\ref{fig.r0} \citep{morioka08,shioya08,nakajima08,geach08}. These 
surveys are small and sampling variance is not always included in the error 
bar quoted on the correlation length (see Orsi et~al. 2008 for an illustration 
of how sampling variance can affect measurements of the correlation function 
made from small fields). The models are in reasonable agreement 
with the estimate by \citet{geach08} at $z=2.2$, but overpredict the low 
redshift measurements. The $z=0.24$ measurements are particularly challenging to 
reproduce. The correlation length of the dark matter in the $\Lambda$CDM model is 
around $5 h^{-1}$Mpc at this redshift, so the $z=0.24$ result implies an 
effective bias of $b<0.5$. Gao \& White (2007) show that dark matter 
haloes at the resolution limit of the Millennium Simulation, 
$M \sim 10^{10}h^{-1}M_{\odot}$, do not reach this level of bias, 
unless the 20\% of the youngest haloes of this mass are selected.  
In the \bowr\ model, the \ha\ emitters populate a range of halo masses, 
with a spread in formation times, and so the effective bias is closer  
to unity. Another possible explanation for the discrepancy is that 
the observational sample could be contaminated by objects which are 
not \ha\ emitters and which dilute the clustering signal. 

The bottom panel of Fig.~\ref{fig.r0} shows the correlation length evolution 
for different H-band selections, compared to observational estimates 
from \citet{firth02}. Note that the samples analysed by Firth et~al. are 
significantly brighter than the typical samples considered in this paper 
($H_{\rm AB} = 20$ versus $H_{\rm AB}=22$). Firth et~al. use photometric 
redshifts to isolate galaxies in redshift bins before measuring the angular 
clustering. Two sets of observational estimates are shown for each magnitude 
limit, corresponding to two choices for the assumed evolution of clustering. 
Again the models display somewhat stronger clustering than the observations 
would suggest at low redshift. The \baur\ model predicts a clustering length 
which increases with redshift. The \bowr\ model, on the other hand, predicts 
a peak in the correlation length around $z \sim 0.7$, with a decline to 
higher redshifts. This reflects the form of the luminosity - halo mass 
relation for galaxy formation models with AGN feedback \citep{kim09}. 
The slope of the luminosity - mass relation changes at the mass for which 
AGN heating becomes important. Coupled with the appreciable scatter in the 
predicted relation, this can result in the brightest galaxies residing in 
haloes of intermediate mass. 

\begin{figure*}
\centering
\includegraphics[width=6.8cm,angle=90]{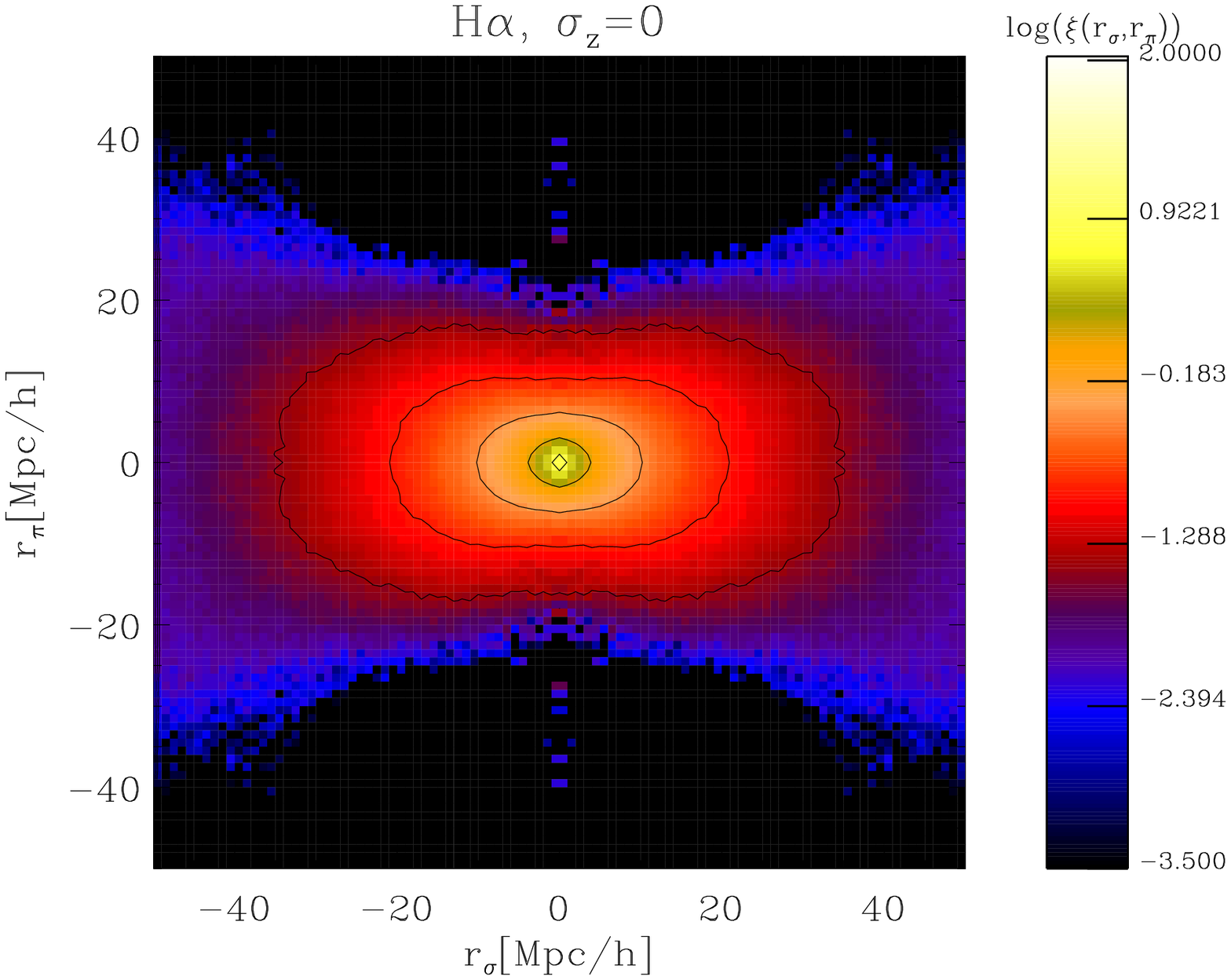}
\includegraphics[width=6.8cm,angle=90]{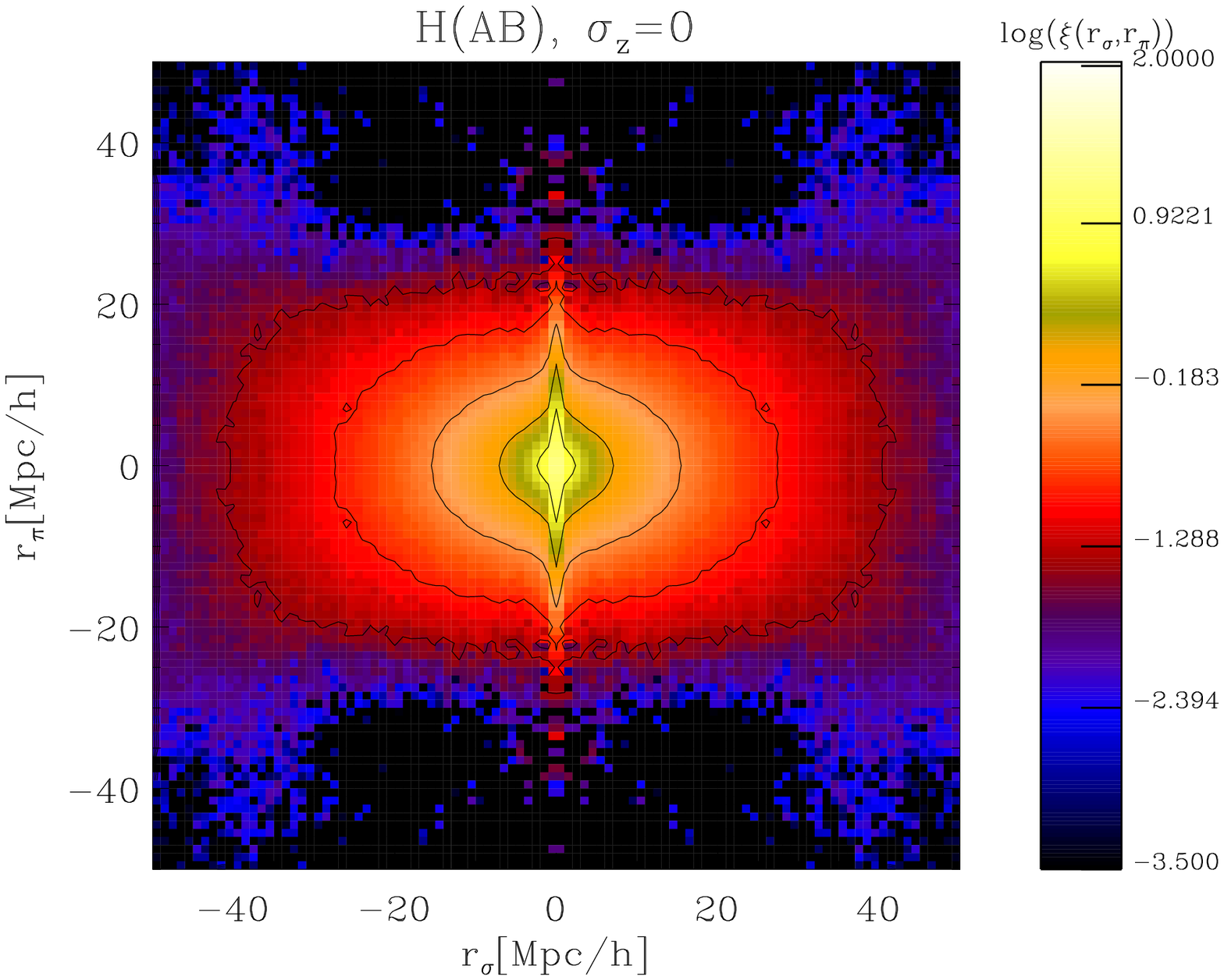}
\includegraphics[width=6.8cm,angle=90]{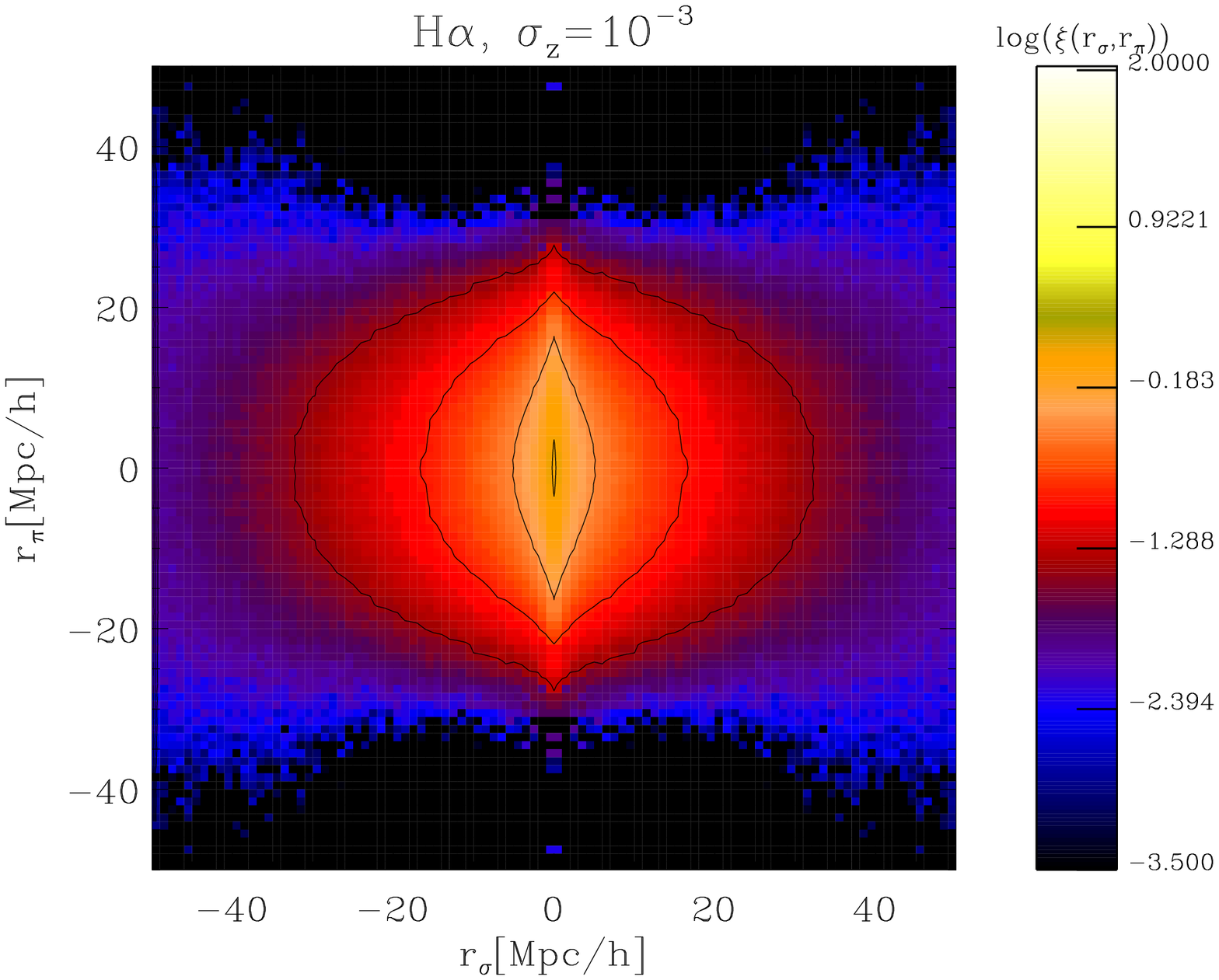}
\includegraphics[width=6.8cm,angle=90]{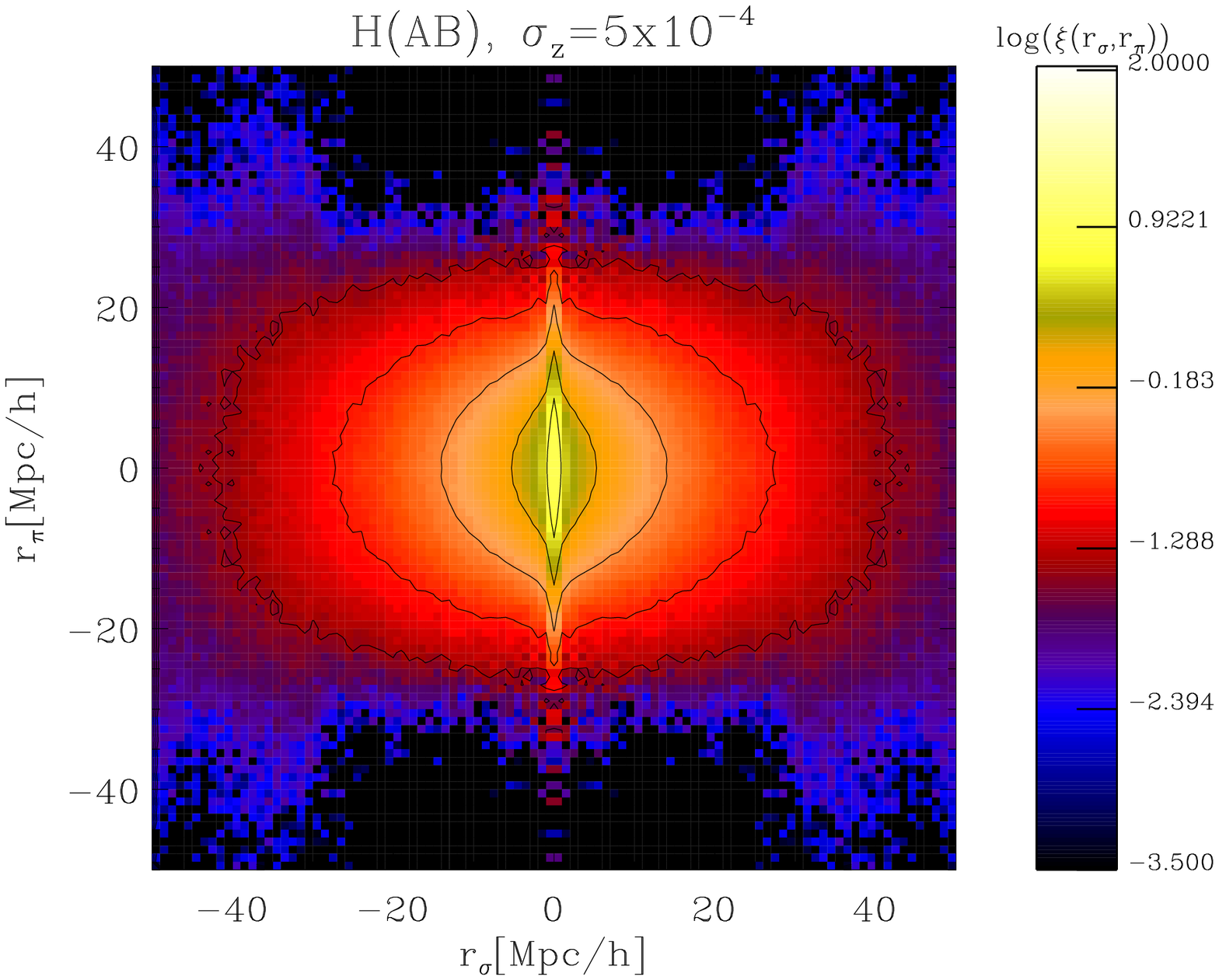}
\includegraphics[width=6.8cm,angle=90]{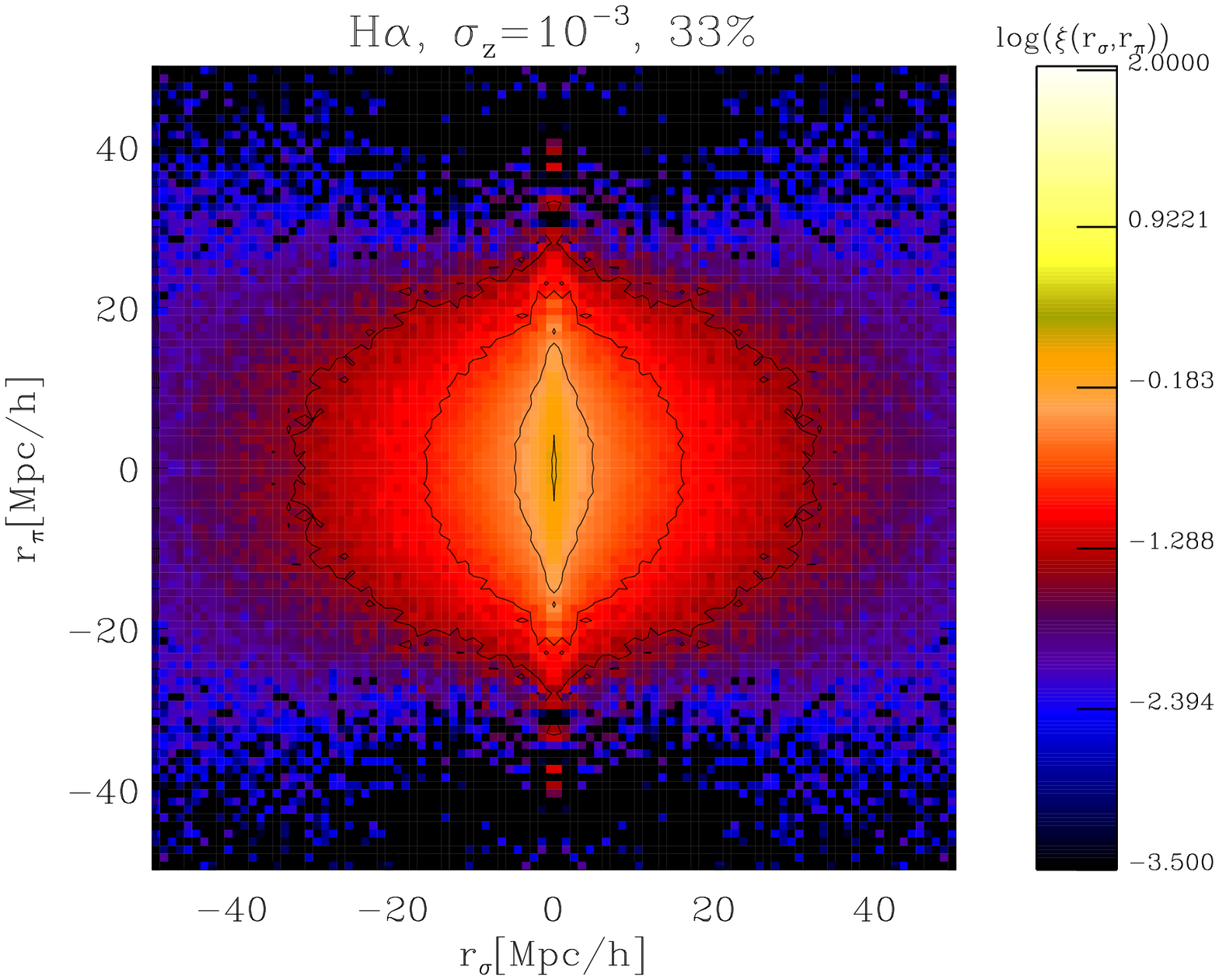}
\includegraphics[width=6.8cm,angle=90]{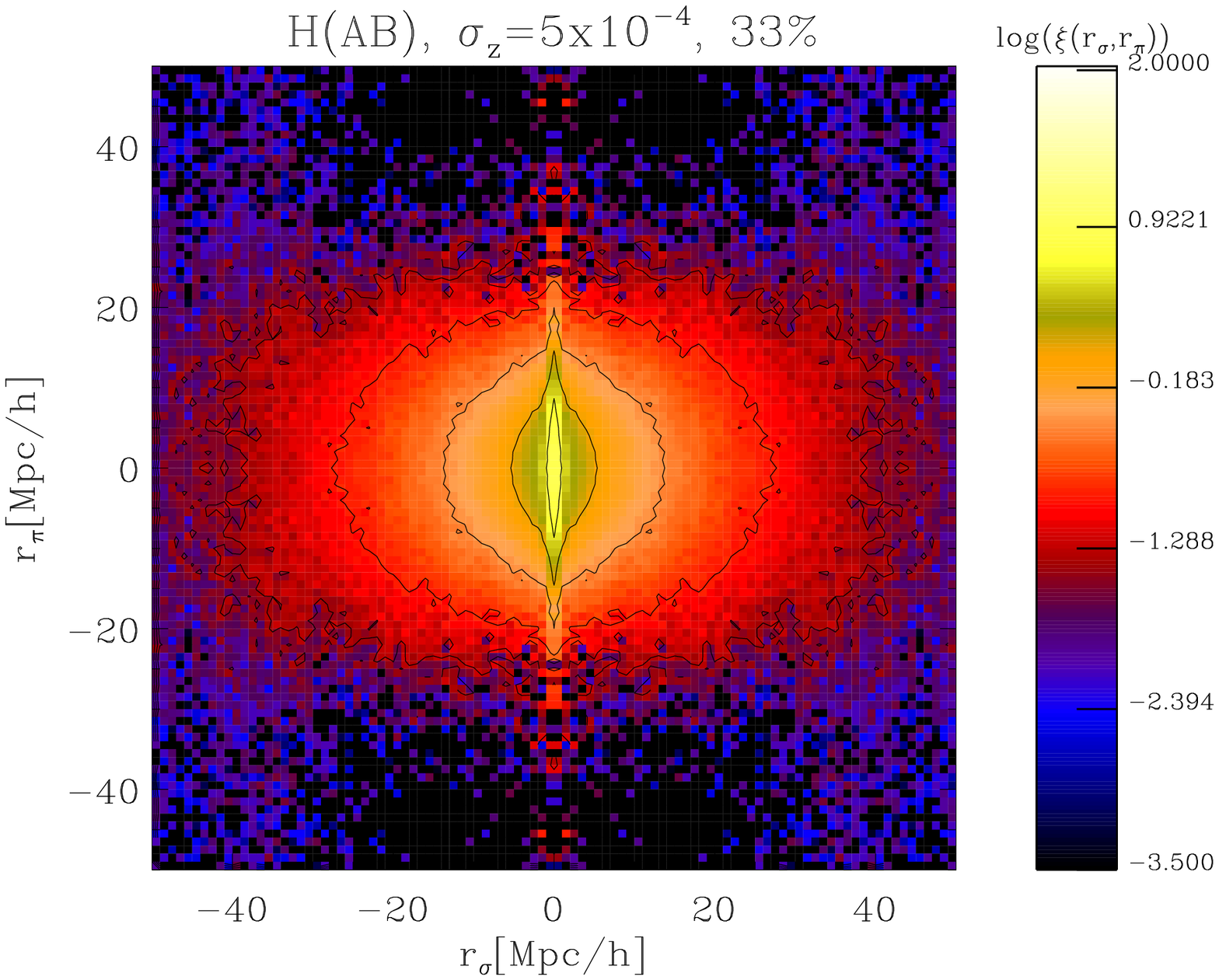}
\caption{
The two point correlation function, measured in redshift space, 
plotted in bins of pair separation parallel ($r_{\pi}$) and perpendicular 
($r_{\sigma}$) to the line of sight, $\xi(r_{\sigma},r_{\pi})$, for \ha\ emitters 
(left-hand panels) and H-band selected (right-hand panels) galaxies in the 
Millennium simulation. The 
samples used are those plotted in Fig.~\ref{fig.dm}. The pair counts 
are replicated over the four quadrants to enhance the visual impression of 
deviations from circular symmetry. The \ha\ catalogue has a limiting flux 
of $\logfha>-16$ and an equivalent width cut of $\ewobs>100$\AA{}; 
the H-band magnitude limit is $H_{\rm AB}=22$. The contours show 
where ${\rm log}(\xid) = [0.5, 0.0, -0.5, -1.0, -1.5]$, from small 
to large pair separations. The upper panels show the correlation 
function measured in fully sampled catalogues without redshift errors. 
The middle panels show how redshift errors change the clustering pattern.  
Representative errors for the two redshift measurements are used: 
$\sigma_z=10^{-3}$ for the slitless case (\ha\ emitters), and 
$\sigma_z=2\times 10^{-4}$ for the slit based measurement 
(H-band selected). 
In the upper and middle panels, all the galaxies are used to compute 
the correlation function. In the bottom panels, only 33\% of the 
galaxies are used in each case, which is indicative of the likely redshift 
success rate for a survey from space. 
}
\label{fig.xi2d}
\end{figure*}

\subsection{Redshift-space distortions}

The amplitude of gravitationally induced bulk flows is sensitive 
to the rate at which perturbations grow, which depends on the expansion 
history of the universe and the nature of the dark energy 
\citep{wang08,guzzo08}. Bulk flows can be measured by their impact on 
the correlation function of galaxies when plotted as a function of 
pair separation perpendicular and parallel to the line of sight, 
$\xi(r_{\sigma}, r_{\pi})$ \citep{hawkins03,ross07}. We now restrict  
our attention to the \bowr\ model, since this is set in the Millennium 
Simulation and we can measure the clustering of the model galaxies directly. 
As the Millennium simulation has periodic boundary conditions, we 
can estimate the correlation function as follows: 
\begin{eqnarray}
 \xid & = & \frac{DD_{\sigma,\pi}}{N\bar{n} \Delta V_{\sigma,\pi}} - 1,\\
\label{eq.xi2d}
\Delta V_{\sigma,\pi} & = & 2\pi r_{\sigma} \Delta r_{\sigma} \Delta r_{\pi},
\label{eq.vol}
\end{eqnarray}
where $DD_{\sigma,\pi}$ is the number of distinct galaxy pairs 
in a bin of pair separation centred on $(r_{\sigma},r_{\pi})$,
$\Delta r_{\sigma}$ and $\Delta r_{\pi}$ are the widths of the 
bins in the $r_{\sigma}$ and $r_{\pi}$ directions, respectively, 
$N$ and $\bar{n}$ are the total number of galaxies and the number 
density of galaxies in the sample, and $\Delta V_{r_{\sigma},r_{\pi}}$ 
corresponds to the volume enclosed in an annulus centred on 
$(r_{\sigma},r_{\pi})$. Note that to avoid any confusion, here 
we refer to the line of sight separation as $r_{\pi}$ and use 
$\pi$ to denote the mathematical constant. 

In redshift surveys, the radial distance to a galaxy is inferred 
from its redshift. The measured redshift contains a contribution 
from the expansion of the Universe, along with a peculiar velocity 
which is induced by inhomogeneties in the density field 
around the galaxy. Thus the position inferred from the redshift 
is not necessarily the true position. The distortion of the clustering 
pattern resulting from peculiar velocities is referred 
to as the redshift space distortion. On large scales, coherent 
motions of galaxies from voids towards overdense regions lead to a 
boost in the clustering amplitude \citep{kaiser87}: 
\begin{equation}
 \frac{\xi(s)}{\xi(r)} = 1 + \frac{2}{3}\beta + \frac{1}{5}\beta^2,
\label{eq.beta1}
\end{equation}
where $\xi(s)$ is the spherically averaged, redshift space 
correlation function, and $\xi(r)$ is its equivalent in real space 
(i.e. without the contribution of peculiar velocities). Eq.~\eqref{eq.beta1}
holds in linear perturbation theory in the distant observer approximation when
gradients in the bulk flow and the effect of the velocity dispersion are small \citep{cole94a,scoccimarro04}.
Strictly speaking, these approximations apply better on large scales.
The parameter $\beta$ is related to the linear growth rate, $D$, through
\begin{eqnarray}
\label{eq.beta2a}
 \beta_{\rm lin} &= &\frac{1}{b} \frac{ {\rm d}\ln D}{ {\rm d}\ln a}, \\
       &\approx &\frac{\om(z)^{\gamma}}{b},
\label{eq.beta2b} 
\end{eqnarray}
where $a$ is the expansion factor. The approximation in 
Eq.~\eqref{eq.beta2b} is valid for an open cosmology, in 
which $\gamma$ is traditionally approximated to $0.6$ \citep{peebles80}.
\citet{lahav91} showed that this approximation should be 
modified in the case of a CDM model with a cosmological 
constant, to display a weak dependence on $\Lambda$. 
\citet{lue04} pointed out that the value of $\gamma$ allows one to 
differentiate between modified gravity and dark energy, since 
$\beta(z) \simeq \Omega_m(a)^{2/3}/b$ for DGP gravity models, while
$\beta(z) \simeq \Omega_m(a)^{5/9}/b$ for a flat Universe with a cosmological
constant.

On small scales, the randomised motions of galaxies inside 
virialised structures lead to a damping of the redshift space 
correlation function and a drop in the ratio $\xi(s)/\xi(r)$\citep{cole94a}.

The impact of peculiar velocities on the clustering of galaxies 
is clearly seen in \xid. The top panels of Fig.~\ref{fig.xi2d} 
show the correlation function of \ha\ emitters selected to have  
$\logfha>-16$ and $\ewobs>100$\AA{} (left) and H-band selected 
galaxies with $H_{\rm AB}<22$ (right). In the top and middle rows 
of Fig.~\ref{fig.xi2d}, all galaxies are used down to the respective 
flux limits.  To obtain clustering in redshift space, we use the 
distant observer approximation and give the galaxies a displacement 
along one of the cartesian axes, as determined by the component 
of the peculiar velocity along the same axis. Without peculiar 
velocities, contours of constant clustering amplitude in \xid would 
be circular. In redshift space, the clustering of H-band selected 
galaxies exhibits a clear signature on small scales of a contribution 
from high velocity dispersion systems -- the so called ``fingers of God''. 
This effect is less evident in the clustering of the \ha\ sample, as  
these galaxies avoid massive haloes, as shown in Fig.~\ref{fig.dm}. 
On large scales, the contours of equal clustering are flattened due 
to coherent flows. Similar distortions have been measured in surveys 
such as the 2dFGRS (\citealt{hawkins03}) and the VLT-VIMOS deep survey 
(\citealt{guzzo08}).

In practice, the measured correlation functions will look somewhat 
different to the idealised results presented in the top row of 
Fig.~\ref{fig.xi2d}. The redshift measurements will have errors, and 
the errors for slitless spectroscopy are expected to be bigger 
than those for slit-based spectroscopy (Euclid-NIS team, 
private communication). We model this by adding a Gaussian distributed 
 velocity, $v_r$, to the peculiar velocities following 
$\delta z = (1+z) v_r /c$. The dispersion of the Gaussian 
is parametrized by  $\sigma_z \equiv \langle \delta z^2\rangle^{1/2}/(1+z)$. 
We show the impact on the predicted clustering of adding 
illustrative redshift uncertainties to the position measurements 
in the middle and bottom panels of Fig.~\ref{fig.xi2d}. For \ha-emitters, 
we chose a fiducial error of $\sigma_z = 10^{-3}$, based on simulations 
by the {\tt Euclid} NIS team. The errors on the slit-based redshifts 
are expected to be at least a factor of 2 times smaller than the 
slitless errors, so we set $\sigma_z = 5\times 10^{-4}$ for the 
$H_{\rm AB}$ selected sample. The impact of the redshift errors 
is most prominent in the case of the \ha\ sample, where the 
contours of constant clustering become more elongated along the 
line-of-sight direction. 

A measure of how well bulk flows can be constrained can be gained 
from the accuracy with which $\beta$ can be measured 
(Eq.~\eqref{eq.beta2a}). We estimate $\beta$ by applying 
Eq.~\eqref{eq.beta1} to the ratio of the redshift space to real 
space correlation function on pair separations between $15 - 30 h^{-1}$Mpc, 
which is close to the maximum pair separation out to which we 
can reliably measure clustering in the Millennium simulation 
volume. The introduction of redshift errors forces us to apply Eq.~\eqref{eq.beta1}
to the measurements from the Millennium simulation on larger scales than 
in the absence of errors. We note that the ratio is noisy even for a box of the
volume of the Millennium, and in practice we average the ratio by projecting
down each of the cartesian axes. The real space correlation function 
is difficult to estimate on large scales, so a less direct approach 
would be applied to actual survey data \citep[see e.g. ][]{guzzo08}. 
Hence, our results will be on the optimistic side of what is likely 
to be attainable with future surveys. Ideally, we would like to apply 
Eq.~\eqref{eq.beta2b} on as large a scale as possible. Kaiser's 
derivation assumes that the perturbations are in the linear regime. 

\begin{table*}
\caption{Values of $\beta$ estimated from the ratio of the redshift space 
to real space correlation function for the fiducial samples at $z=1$.
We consider \ha\ emitters with fluxes $\logfha > [-15.4,-16]$ and 
H-band selected galaxies with $H_{AB}<[22,23]$. The table is divided 
into two parts. The first half assumes a redshift success rate of $100\%$ 
and the second a $33\%$ redshift success rate. Each segment is divided 
into two, showing the impact on $\beta$ of including the expected redshift 
uncertainties: $\sigma_z = 10^{-3}$ for \ha\ emitters and 
$\sigma_z = 5\times 10^{-4}$ for H-band selected samples. Column (1) shows 
$\beta_{\rm lin}$, the exact theoretical value of $\beta$ obtained 
when using Eq.~\eqref{eq.beta2a}.
Column (2) shows $\beta_m$, the value of $\beta$ measured in the 
simulation including the 1 $\sigma$ error.
Column (3) shows the fractional error on $\beta_m$ using the Millennium volume.
Column (4) shows the fractional error on $\beta_m$
obtained when using mock catalogues from the BASICC simulation.}
\label{table.beta}

\begin{tabular}{@{}lcccc}
\hline
\hline
    & (1)  &   (2) & (3)  &  (4)\\
\hline
   &  $\beta_{lin}$ & $\beta_m$ & $(\delta \beta_m / \beta_{lin}) $ &  $(\delta \beta_m /\beta_{lin})  $ \\
 & & & Millennium & BASICC \\
\hline
Sampling rate = 100\% & & & &\\
\hline
$\log(F(H\alpha))>-15.4$, $\sigma_z =0$ & $0.761$ & $0.684\pm 0.153$ & $0.201$  & $0.125$ \\
$\log(F(H\alpha))>-16.0$, $\sigma_z =0$ & $0.821$ & $0.766\pm 0.027$ & $0.034$  & $0.021$ \\
$H(AB)<22$, $\sigma_z =0$ & $0.521$ & $0.491\pm 0.026$ & $0.051$  & $0.019$ \\
$H(AB)<23$, $\sigma_z =0$ & $0.565$ & $0.536\pm 0.013$ & $0.023$  & $0.013$ \\
\hline
$\log(F(H\alpha))>-15.4$, $\sigma_z =10^{-3}$ & $0.761$ & $0.768\pm 0.170$ & $0.224$  & $0.122$ \\
$\log(F(H\alpha))>-16.0$, $\sigma_z =10^{-3}$ & $0.821$ & $0.825\pm 0.058$ & $0.071$  & $0.081$ \\
$H(AB)<22$, $\sigma_z =5\times 10^{-4}$ & $0.521$ & $0.527\pm 0.029$ & $0.057$  & $0.012$ \\
$H(AB)<23$, $\sigma_z =5\times 10^{-4}$ & $0.565$ & $0.569\pm 0.012$ & $0.022$  & $0.008$ \\
\hline
\hline
Sampling rate = 33\% & & & &\\
\hline
$\log(F(H\alpha))>-15.4$, $\sigma_z =0$ & $0.634$ & $0.123\pm 0.447$ & $0.704$  & $0.449$ \\
$\log(F(H\alpha))>-16.0$, $\sigma_z =0$ & $0.807$ & $0.680\pm 0.104$ & $0.129$  & $0.033$ \\
$H(AB)<22$, $\sigma_z =0$ & $0.516$ & $0.482\pm 0.049$ & $0.095$  & $0.036$ \\
$H(AB)<23$, $\sigma_z =0$ & $0.568$ & $0.569\pm 0.029$ & $0.051$  & $0.018$ \\
\hline
$\log(F(H\alpha))>-15.4$, $\sigma_z =10^{-3}$ & $0.634$ & $0.300\pm 0.216$ & $0.341$  & $0.341$ \\
$\log(F(H\alpha))>-16.0$, $\sigma_z =10^{-3}$ & $0.807$ & $0.749\pm 0.118$ & $0.146$  & $0.078$ \\
$H(AB)<22$, $\sigma_z =5\times 10^{-4}$ & $0.516$ & $0.494\pm 0.061$ & $0.118$  & $0.023$ \\
$H(AB)<23$, $\sigma_z =5\times 10^{-4}$ & $0.568$ & $0.603\pm 0.028$ & $0.050$  & $0.012$ \\
\hline
\hline
\end{tabular}

\end{table*}

We solve the integral for the growth rate $D$ in Eq.~\eqref{eq.beta2a} 
\citep[see ][]{lahav91} and use this exact result with the value of the 
bias $b$ measured for each galaxy sample to get the theoretical 
value $\beta_{\rm lin}$.  
Table~\ref{table.beta} shows the comparison between $\beta_m$, the measured value 
of $\beta$ in the simulation, and target theoretical value $\beta_{\rm lin}$. Two different 
selection cuts are chosen for both \ha\ and H-band samples to cover a range of 
survey configurations: $\logfha>[-15.4,-16.0]$ for \ha\ samples and 
$H_{AB}<[22,23]$ for the magnitude limited samples. All the mock catalogues 
studied return a value for $\beta_m$ which is systematically below 
the expected value, $\beta_{\rm lin}$.

\begin{figure}
 \centering
\includegraphics[width=11cm,angle=90]{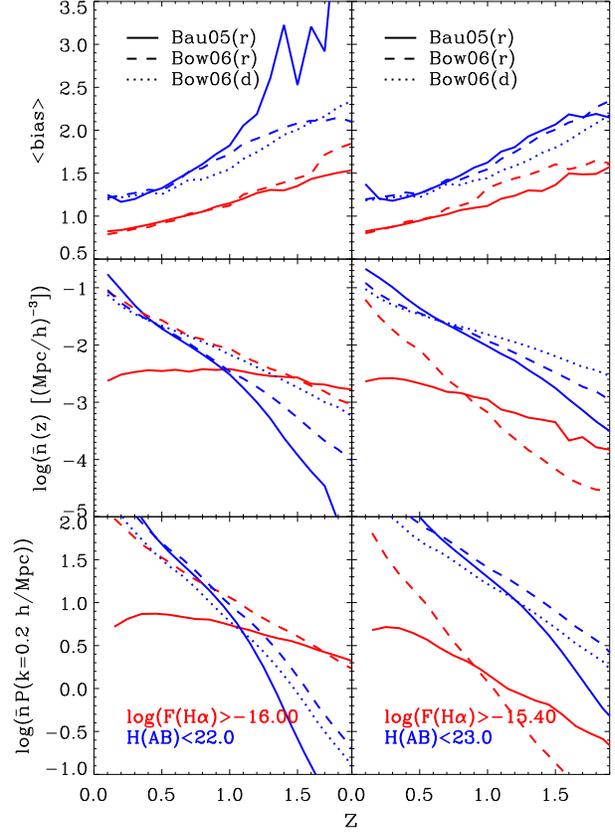}
\caption{
The effective bias (top panel), number density of galaxies 
(middle panel) and the product $\bar{n}P$ (bottom panel) as functions of redshift, where $P$ is measured
at wavenumber $k = 0.2$ Mpc/h. The solid lines show the predictions for the \baur\ model and the 
\bowr\ model is shown using dashed lines. The two columns show different \ha\ and H-band selections: 
In the first column the \ha\ sample is defined 
by a limiting flux of $\logfha>-16$ and $\ewobs>100$\AA{} (red curves). 
The magnitude limited sample has $H_{AB}<22$ (blue curves).
In the second column the \ha\ sample has $\logfha>-15.4$ and $\ewobs>100$\AA{}, 
and the H-band sample has $H(AB)<23$. In all panels 
the redshift success rate considered is 100\%. 
}
\label{fig.fiducial}
\end{figure}

When redshift errors are omitted and a 100\% redshift success rate is used, 
both selection methods seem to reproduce the expected value of 
$\beta_{\rm lin}$ to within better than $\sim 10\%$. When redshift errors 
are included, the spatial distribution along the line of sight appears 
more elongated than it would be if the true galaxy positions could be 
used. This leads to an increase in the small scale damping of the 
clustering. However, at the same time contours of constant clustering 
amplitude are pushed out to larger pair separations in the radial direction. 
This results in an increase in the ratio of redshift space to real space clustering 
and an increase in the recovered value of $\beta$.
When including the likely redshift errors, the values of $\beta_m$ found 
are slightly higher than those without redshift errors. This small boost 
in the value of $\beta_m$ is greatest in the \ha\ sample, because of the 
larger redshift errors than in the H-band sample. 

We have also tested the impact of applying different redshift success rates on 
the determination of $\beta_m$. The lower part of Table~\ref{table.beta} shows 
the impact of a $33\%$ redshift success rate. For $\logfha>-15.4$, 
our results for $\beta_m$ shows that it is unlikely to get a robust estimate of 
$\beta$ at this flux limit, because the smaller number density makes the correlation 
functions very noisy, thus making $\beta_m$ impossible to be measured correctly.
In contrast, the impact of a $33\%$ of success rate in the $\logfha>-16$ sample is 
negligible. The $\beta_m$ values calculated using the H-band catalogues 
are also mostly unaffected. When redshift uncertainties are considered, as before, 
the $\beta_m$ values are closer to the theoretical $\beta_{lin}$. 
Hence redshift uncertainties will contribute to the uncertainty on $\beta_m$, 
but they still permit an accurate determination of $\beta$, provided they 
do not exceed $\sigma_z=10^{-3}$.

The noisy correlation functions for the configurations
with  $\logfha>-15.4$ and sampling rate of 33\% produce measurements of $\beta_m$ with large errors.
The mock catalogues used so far in this section were 
created from the Millennium simulation, which has $V_{\rm Mill} = 500^3 {\rm [Mpc/h]^3}$.
This volume is almost three orders of magnitude smaller than the volume expected in a large 
redshift survey from a space mission like \euclid (see next section). 
In order to test the impact of using this limited volume when measuring $\beta_m$ and its error, 
we plant the \bowr\ model into a larger volume using the BASICC N-body simulation \citep{angulo08a}, 
which has a volume almost 20 times larger than the Millennium run
($V_{\rm BASICC} =  1340^3 {\rm [Mpc/h]^3}$). 
The errors on $\beta_m$ shown in Table ~\ref{table.beta} are expected, to
first order, to scale with the error on the power spectrum (see Eq.~\eqref{eq.dp} below).
If we compare two galaxy samples with the same number density but in different volumes,
then the error on $\beta_m$
should scale as $\delta \beta \propto 1/\sqrt{V}$, where $V$ is the 
volume of the sample.

The only drawback of using the BASICC simulation is 
that the mass resolution is worse than in the Millennium simulation. Haloes with mass greater than $5.5\times 10^{11} M_{\odot}/h$ can be resolved in the BASICC simulation. 
The galaxy samples studied here are hosted by haloes with masses greater than $\sim 8\times 10^{10} M_{\odot}/h$, 
so if we only plant galaxies into haloes resolved in the BASICC run then we would miss a substantial fraction of the galaxies. 
To avoid this incompleteness, those galaxies which should be hosted by haloes below the mass resolution limit are planted 
on randomly selected ungrouped particles, i.e. dark matter particles which do not belong to any halo. 
This scheme is approximate and works best if the unresolved haloes have a bias close to unity, i.e. where the
bias is not a strong function of mass. This is almost the case in the application of this method to the BASICC run, so the 
clustering amplitude appears slightly boosted for all the configurations studied here. However, 
since we only want to study the variation in the error on $\beta_m$ when using a larger volume, 
we apply the same method described above to measure $\beta_m$ in the galaxy samples planted in the BASICC run. \\

As shown in the fourth column of Table ~\ref{table.beta}, we find that for all the \ha\ configurations here 
studied the error on $\beta_m$ obtained when using the BASICC simulation is a factor 1-6 smaller 
than that found with the Millennium samples. 
The H-band samples, on the other hand, have errors roughly $\sim 4$ times smaller in the BASICC volume compared to the Millennium volume, which is what we expect if we assume that the error on $\beta_m$ scales with $1/\sqrt{V}$. 

The \euclid\ survey will cover a geometrical volume of $\sim 90 {\rm [Gpc/h]^3}$ with an effective volume of around half of this (see next section). We expect
that \euclid\ should meausre $\beta_m$ with an accuracy around 5 times smaller than that estimated for the galaxy samples planted into the BASICC simulation.


\subsection{Effective survey volume}
\label{sec.veff}

Ongoing and future surveys aim to measure the baryonic acoustic 
oscillation (BAO) signal in the power spectrum of galaxies. 
The primary consideration for an accurate power spectrum measurement 
is to maximize the survey 
volume in order to maximize the number of independent $k$-modes. 
However, because the power spectrum is measured using a finite 
number of galaxies there is an associated discreteness noise. 
The number density of galaxies in a flux limited sample drops 
rapidly with increasing redshift, which means that discreteness 
noise also increases. When the discreteness noise becomes comparable 
to the power spectrum amplitude, it is difficult to measure the 
clustering signal. This trend is encapsulated in the expression for 
the fractional  error on the power spectrum derived by Feldman, Kaiser \& 
Peacock (1994): 
\begin{eqnarray}
\label{eq.dp}
 \frac{\sigma}{P} & \approx &  \frac{2\pi}{\sqrt{Vk^2\Delta k}}\left(1 + \frac{1}{\bar{n}P}\right), \\
  & \approx & \frac{2 \pi}{\sqrt{V_{\rm eff}(k) k^{2} \Delta k}}, 
\end{eqnarray}
where $\sigma$ is the error on the power spectrum $P$, $V$ is the 
geometrical survey volume and $\bar{n}$ is the number density of 
galaxies.  When the contrast of the power spectrum is high, i.e. 
$\bar{n} P \gg 1$, then the fractional error scales as the inverse 
square root of the survey volume. However, in the case that 
$\bar{n} P \le 1$, the gain in accuracy from increasing the survey 
volume is less than the inverse square root of the 
increased volume. The amplitude of the power spectrum compared 
to the discreteness noise of the galaxies used to trace the density 
field is therefore a key consideration when assessing the effectiveness 
of different tracers of the large scale structure of the Universe. 

\begin{figure*}
\centering
\includegraphics[width=15cm,angle=90]{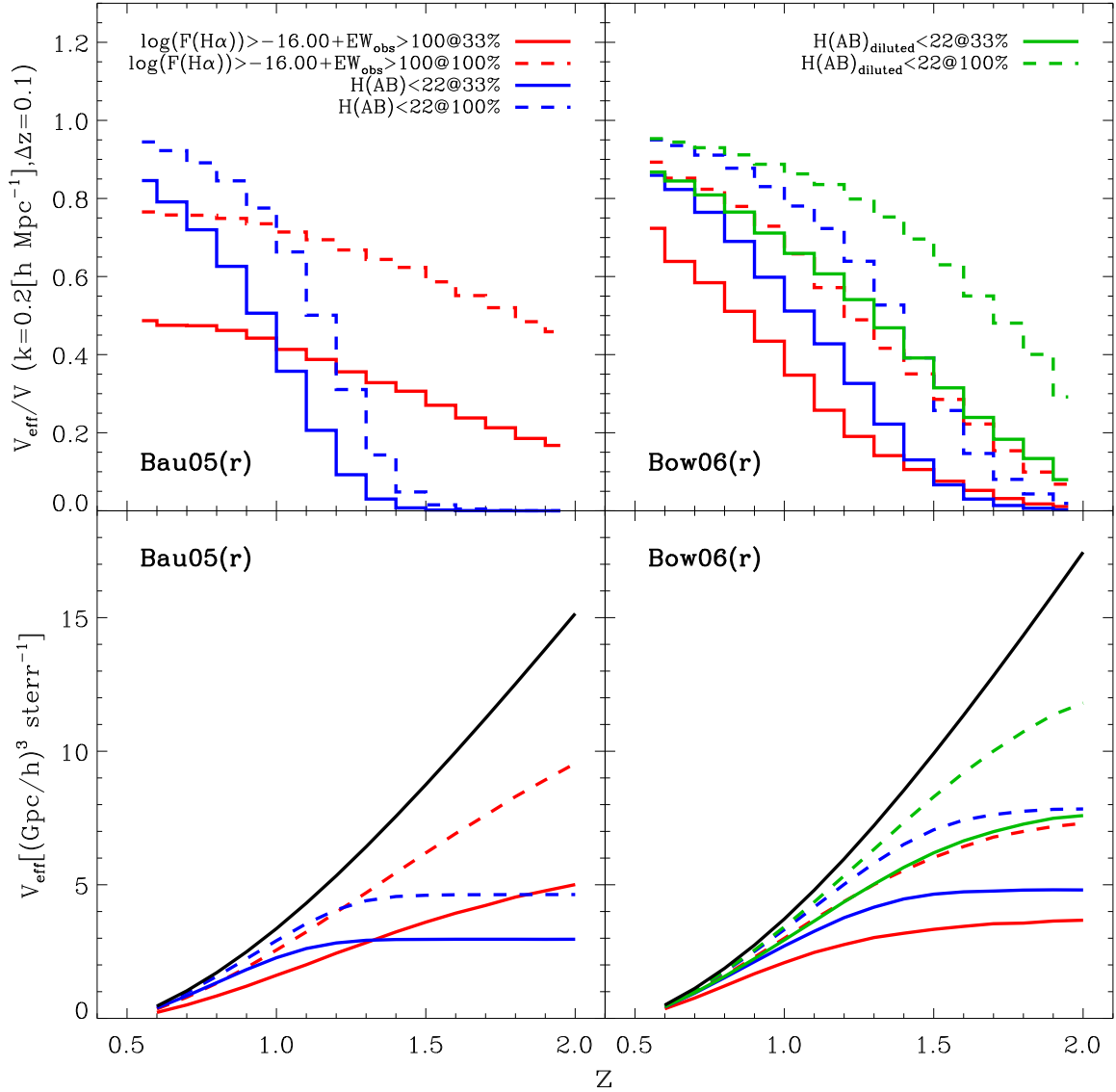}
\caption{
The effective volume of \ha- and H-band selected samples. 
The left-hand panels show results for \baur\ model and the 
right-hand panels show the \bowr\ model; in the latter case, 
the effective volume for a randomly diluted sample of galaxies from 
the original \bow\ model is also shown. 
The upper row shows the effective volume divided by the geometrical 
volume in redshift shells of width $\Delta z = 0.1$; the power 
spectrum at $k=0.2 h {\rm Mpc}^{-1}$ is used to compute the 
effective volume (see text). The lower panels show the cumulative 
effective volume per steradian starting from $z=0.5$ and extending 
up to the redshift at which the curve is plotted. 
Red curves show the results for \ha\ selected galaxies with 
$\logfha > -16$ and $\ewobs > 100$\AA{}. 
The solid red line shows the result of applying a redshift success of 
$33\%$, whereas the red dashed line assumes a $100\%$ success rate. 
The blue lines show the results for an H-band magnitude selected 
survey with $H_{\rm AB} < 22$. As before, the solid blue line shows 
the results for a sampling rate of $33\%$, and the dashed line 
assumes $100\%$ sampling. The green lines show the results using 
the \bow\ model diluted ($\bowd$) to match the observed number counts; 
as before solid and dashed show 33\% and 100\% success rates, 
respectively. The black solid curves in the bottom panels show the 
total comoving volume covering the redshift range shown.}
\label{fig.veff}
\end{figure*}

\galform\ gives us all the information required to 
estimate the effective volume of a survey with a given selection 
criteria (which defines the number density of galaxies, $\bar n(z)$, and 
the effective bias as a function of redshift). For simplicity, we 
use the linear theory power spectrum of dark matter, which is a 
reasonable approximation on the wavenumber scales studied here. 
The galaxy power spectrum is assumed to be given by $P_{\rm g}(k,z) 
= b(z)^2 P_{\rm dm}(k,z)$, where $b(z)$ is the effective bias of the 
galaxy sample. We calculate the fraction of volume utilized in a given redshift 
interval following Tegmark (1997),
\begin{equation}
\veff (k, z)  = 
\int_{z_{\rm min}}^{z_{\rm max}} \left[ \frac{\bar{n}(z)P_{\rm g}(k,z)}{1 + \bar{n}(z)P_{\rm g}(k,z)}\right]^2 \frac{{\rm d}V}{{\rm d}z}
{\rm d}z,
\label{eq.veff}
\end{equation}
where all quantities are expressed in comoving coordinates. 
We calculate $\veff/V$ for a range of possible survey configurations 
considering different limits in flux, \ewobs, magnitude limit and 
redshift success rate (see Table~\ref{table.veff}). The redshift range 
is chosen to match that expected to be set by the near-infrared 
instrumentation to be used in future surveys.  

Fig.~\ref{fig.fiducial} shows the predictions from \galform\ 
which are required to compute the effective volume, for two illustrative \ha\ and H-band 
selected surveys, covering the current expected flux/magnitude limits of space missions. 
The bias predicted for H-band galaxies is at least
$\sim 30\%$ higher than that for \ha-emitters in both panels of Fig. \ref{fig.fiducial}.
 This reflects the different spatial 
distribution of these samples apparent in Fig.~\ref{fig.dm}, in 
which is it clear that \ha\ emitters avoid cluster-mass dark matter haloes.  The middle panel of Fig.~\ref{fig.fiducial} 
shows the galaxy number density as a function of redshift for these illustrative 
surveys. For the \ha\ selection, the models predict very different number densities at 
low redshifts, as shown also in Fig~\ref{fig.nz}.
For $z>1$ the \bowr\ model predicts progressively more galaxies than 
the \baur\ model for the H-band selection. Overall, the number density 
of galaxies in the H-band sample at high redshift is much lower than that 
of \ha\ emitters. However, we remind the reader than these scaled models 
match the H-band counts but have a shallower redshift distribution than is 
suggested by the observations. The bottom panel of Fig.~\ref{fig.fiducial} 
shows the power spectrum times the shot noise, $\bar{n}P$, as a function of 
redshift. A survey which efficiently samples the available volume will 
have $\bar{n}P >1$. The slow decline of the number density of \ha\ galaxies 
with redshift in the \baur\ model is reflected in $\bar{n}P>1$ throughout the 
redshift range considered here, whereas in the \bowr\ model, the \ha\ sample 
has a very steeply falling $\bar{n}P$ curve, with $\bar{n}P<1$ for $z>1.5$. 
The predictions of $\bar{n}P$ for the H-band are similar in both models, 
dropping below 1 at $z\sim 1.3-1.5$. 

The predictions for the bias, number density and power spectrum of galaxies plotted 
in Fig.~\ref{fig.fiducial} are used in Eq.~\eqref{eq.veff} to 
calculate the effective volume, which is shown in Fig.~\ref{fig.veff}. 
The top panels show the differential $\veff/V$ calculated in shells 
of $\Delta z = 0.1$ for redshifts spanning the range $z=[0.5,2]$. 
The bottom panels of Fig.~\ref{fig.veff} show the cumulative \veff\ 
contained in the redshift range from $z=0.5$  up to $z=2$. 
We follow previous work and use the amplitude of the power spectrum 
at $k=0.2 h {\rm Mpc^{-1}}$, which roughly corresponds to the centre 
of the wavenumber range over which the BAO signal is measured. 
We show the result for the fiducial survey selections with different 
redshift success rates, 100\% and 33\%. In addition, for the H-band selected 
survey, we also show the results obtained with the alternative approach 
discussed in the previous section, in which the galaxies in the 
\bow\ sample are diluted by a factor of 0.63.

\begin{table*}
\caption{
The effective volume of \ha- and H-band selected surveys for different 
selection criteria. We evaluate a given survey configuration 
in terms of its effective volume in the redshift range $0<z<2$ (top) 
and $0.5<z<2$ (bottom), which is expressed as a fraction of the geometrical 
volume over the same redshift interval. The first column shows the 
galaxy selection method used, \ha\ for an \ha\ selected survey with a minimum 
flux limit and \ewobs\ cut or $H_{\rm AB}$ for an H-band magnitude 
limited survey. The second column shows the H-band magnitude limit 
chosen in a given configuration, where applicable. The third column 
shows the minimum \ha\ flux chosen, again where applicable, and the 
fourth column the minimum \ewobs cut applied. The fifth column shows 
the redshift success rate assumed. Columns 6, 7 and 8 show the fractional 
effective volume obtained for a given configuration in the \bau, \bow\ and 
the diluted version of the \bow\  model respectively. Finally, 
columns 9, 10 and 11 show our estimate of the corresponding 
percentage error on the determination of $w$, the dark energy 
equation of state parameter, for the \bau, \bow\ and 
diluted \bow\ models, respectively.
}
\label{table.veff}
\begin{tabular}{@{}lccccccccccc}
\hline
\hline
         Selection & $H_{\rm AB}$  & ${\rm log(F_{H\alpha})}$ & $\ewobs$ & Sampling & $\veff/V$ & $\veff/V$ & $\veff/V$ & $\Delta w (\%)$ & $\Delta w (\%)$&  $\Delta w (\%)$\\
		    & (mags) & ${\rm (erg s^{-1}cm^{-2})}$ & (\AA{}) & 	rate &\baur	&\bowr & $\bow_{\rm dil}$ &\baur	&\bowr & $\bow_{\rm dil}$ \\
\hline
\hline
$0<z<2$ &           &        &          &         &              &                 &                 \\
\hline
\hline  
\ha\ & - & -15.40 & 100 & 0.33 & 0.08 & 0.09 & 0.00& 1.2 & 1.1 &0.0 \\
\ha\ & - & -15.40 & 100 & 1.00 & 0.24 & 0.18 & 0.00& 0.7 & 0.8 &0.0 \\
\ha\ & - & -15.40 & 0 & 1.00 & 0.24 & 0.18 & 0.00& 0.7 & 0.8 &0.0 \\
\ha\ & - & -15.70 & 100 & 0.33 & 0.19 & 0.20 & 0.00& 0.8 & 0.7 &0.0 \\
\ha\ & - & -15.70 & 100 & 1.00 & 0.44 & 0.39 & 0.00& 0.5 & 0.5 &0.0 \\
\ha\ & - & -15.70 & 0 & 1.00 & 0.45 & 0.39 & 0.00& 0.5 & 0.5 &0.0 \\
\ha\ & - & -16.00 & 100 & 0.33 & 0.34 & 0.41 & 0.00& 0.6 & 0.5 &0.0 \\
\ha\ & - & -16.00 & 100 & 1.00 & 0.63 & 0.67 & 0.00& 0.4 & 0.4 &0.0 \\
\ha\ & - & -16.00 & 0 & 1.00 & 0.64 & 0.67 & 0.00& 0.4 & 0.4 &0.0 \\
H(AB) & 21 & - & - & 0.33 & 0.13 & 0.13 & 0.22& 1.0 & 0.9 &0.7 \\
H(AB) & 21 & - & - & 1.00 & 0.18 & 0.21 & 0.38& 0.8 & 0.7 &0.5 \\
H(AB) & 22 & - & - & 0.33 & 0.23 & 0.30 & 0.45& 0.7 & 0.6 &0.5 \\
H(AB) & 22 & - & - & 1.00 & 0.33 & 0.47 & 0.68& 0.6 & 0.5 &0.4 \\
H(AB) & 23 & - & - & 0.33 & 0.41 & 0.57 & 0.68& 0.5 & 0.4 &0.4 \\
H(AB) & 23 & - & - & 1.00 & 0.59 & 0.78 & 0.86& 0.4 & 0.3 &0.3 \\ 
&0.3 \\
\hline
\hline
$0.5<z<2$ &	&	&	&	&	&	&	&	\\
\hline
\hline
\ha\ & - & -15.40 & 100 & 0.33 & 0.06 & 0.06 & 0.00& 1.4 & 1.4 &0.0 \\
\ha\ & - & -15.40 & 100 & 1.00 & 0.21 & 0.15 & 0.00& 0.8 & 0.9 &0.0 \\
\ha\ & - & -15.40 & 0 & 1.00 & 0.21 & 0.15 & 0.00& 0.8 & 0.9 &0.0 \\
\ha\ & - & -15.70 & 100 & 0.33 & 0.18 & 0.07 & 0.00& 0.9 & 1.2 &0.0 \\
\ha\ & - & -15.70 & 100 & 1.00 & 0.43 & 0.17 & 0.00& 0.5 & 0.8 &0.0 \\
\ha\ & - & -15.70 & 0 & 1.00 & 0.44 & 0.17 & 0.00& 0.5 & 0.8 &0.0 \\
\ha\ & - & -16.00 & 100 & 0.33 & 0.33 & 0.21 & 0.00& 0.6 & 0.7 &0.0 \\
\ha\ & - & -16.00 & 100 & 1.00 & 0.62 & 0.41 & 0.00& 0.4 & 0.5 &0.0 \\
\ha\ & - & -16.00 & 0 & 1.00 & 0.63 & 0.41 & 0.00& 0.4 & 0.5 &0.0 \\
H(AB) & 21 & - & - & 0.33 & 0.09 & 0.10 & 0.19& 1.2 & 1.1 &0.8 \\
H(AB) & 21 & - & - & 1.00 & 0.14 & 0.18 & 0.35& 1.0 & 0.8 &0.6 \\
H(AB) & 22 & - & - & 0.33 & 0.19 & 0.27 & 0.43& 0.8 & 0.6 &0.5 \\
H(AB) & 22 & - & - & 1.00 & 0.30 & 0.44 & 0.67& 0.6 & 0.5 &0.4 \\
H(AB) & 23 & - & - & 0.33 & 0.38 & 0.55 & 0.67& 0.6 & 0.4 &0.4 \\
H(AB) & 23 & - & - & 1.00 & 0.57 & 0.77 & 0.86& 0.5 & 0.4 &0.3 \\
\hline
\hline
\end{tabular}
\end{table*}

In general, the effective volume is close to the geometrical volume 
at low redshifts. This is because $\bar{n}P \gg 1$ at these redshifts. 
In the top panels of Fig.~\ref{fig.veff}, where the differential 
\veff/V is plotted in shells of $\Delta z =0.1$,  we see that 
shells at higher redshifts cover progressively smaller differential 
effective volumes. This is due to the overall decrease in the number 
density of galaxies beyond the peak in the redshift distribution 
(see Figs.~\ref{fig.nzhband}, \ref{fig.nz} and \ref{fig.fiducial}), which 
wins out over the more modest increase in the bias of the galaxies 
picked up with increasing redshift. The bottom panels of Fig.~\ref{fig.veff} 
show the same effect: at higher redshifts, the gain in effective volume 
is much smaller than the corresponding gain in the geometrical volume 
of the survey. We remind the reader that our calculation for the 
effective volume in the H-band using models with rescaled luminosities is 
likely to be an underestimate, as these models underpredict the observed 
high redshift tail of the redshift distribution. A better estimate is likely 
to be provided by the \bowd\ model, in which the number of galaxies is 
adjusted by a making a random sampling, rather than by changing their luminosities. 
This case is shown by the green curves in Fig.~\ref{fig.veff}. 

The calculations presented in Fig.~\ref{fig.veff} are extended to 
a range of survey specifications in Table~\ref{table.veff}. 
This table shows calculations for two different redshift ranges: 
$0<z<2$ and $0.5<z<2$, and includes also the effect of applying 
different selection criteria and redshift success rates to \ha\ 
and H-band surveys. An \ha\ survey with a limiting flux of $\logfha >-15.4$, 
an equivalent width $\ewobs>100$\AA{} and a sampling rate of $0.33$, 
similar to the baseline spectroscopic solution for \euclid, would 
have a very small $\veff/V \sim 0.04$ for the redshift interval $z=0.5-2$. 
In contrast, an H-band survey with $H_{\rm AB}<22$ and a sampling 
rate of $0.33$, an alternative spectroscopic solution for \euclid, 
has $\veff/V = 0.19-0.27$ or even up to $\veff/V=0.43$ in the case 
of the diluted model. To reach a comparable effective volume,  a \ha\ survey 
would need to reach a flux limit of at least $\logfha >-16$ (at the same equivalent 
width cut and redshift success rate).

The calculation of the effective volume also allows us to  
make an indicative estimate of the accuracy with which the 
dark energy equation of state parameter, $w$, can be measured 
for a given survey configuration. Angulo et~al. (2008a) used large 
volume N-body simulations combined with the {\tt GALFORM} model 
to calculate the accuracy with which the equation of state 
parameter $w$ can be measured for different galaxy samples. 
They found a small difference ($\sim 10\%$) in the accuracy 
with which $w$ can be measured for a continuum magnitude limited sample and 
an emission line sample with the same number density of objects. 
Their results can be summarised by: 
\begin{equation}
 \Delta w (\%) = \frac{1.5\%}{\sqrt{\veff}},
\label{eq.alpha_veff}
\end{equation}
where \veff is in units of $h^{-3}{\rm Gpc}^{3}$ and the constant 
of proportionality (in this case, 1.5) depends on 
which cosmological parameters are held fixed; in the present case models are 
considered in which the distance to the epoch of last scattering is fixed as 
the dark energy equation of state parameter varies. 
We obtain an estimate of the accuracy with which $w$ can be measured 
by inserting \veff into Eq.~\ref{eq.alpha_veff}, which is shown 
in Table \ref{table.veff}, for the \bau and \bow\ models. 

\section{Discussion and Conclusions}
\label{sec.conclusions}

In this paper we have presented the first predictions for clustering 
measurements expected from future space-based surveys to be conducted with instrumentation 
sensitive in the near-infrared. We have used published galaxy formation 
models to predict the abundance and clustering of galaxies selected by 
either their \ha\ line emission or H-band continuum magnitude. 
The motivation for this exercise is to assess the relative performance 
of the spectroscopic solutions proposed for galaxy surveys 
in forthcoming space missions which have the primary 
aim of constraining the nature of dark energy.  

The physical processes behind \ha\ and H-band emission are quite 
different. \ha\ emission is sensitive to the instantaneous star 
formation rate in a galaxy, as the line emission is driven by 
the number of Lyman continuum photons produced by massive young stars. 
Emission in the observer frame H-band typically probes the rest frame 
$R$-band for the proposed magnitude limits and is
more sensitive to the stellar mass of the galaxy than to 
the instantaneous star formation rate.  

The {\tt GALFORM} code predicts the star formation histories of a wide 
population of galaxies, and so naturally predicts their star formation 
rates and stellar masses at the time of observation. Variation in 
galaxy properties is driven by the mass and formation history of the 
host dark matter halo. This is because the strength of a range of 
physical effects depend on halo properties such as the depth 
of the gravitational potential well or the gas cooling time. 
This point is most 
striking in our plot of the spatial distribution of \ha\ and H-band 
selected galaxies, Fig.~\ref{fig.dm}. This figure shows remarkable 
differences in the way that these galaxies trace the underlying 
dark matter distribution. \ha\ emitters avoid the most massive 
dark matter haloes and trace out the filamentary structures 
surrounding them. The H-band emitters, on the other hand, are 
preferentially found in the most massive haloes. This difference in 
the spatial distribution of these tracers has important consequences 
for the redshift space distortion of clustering. 

In this paper we have studied two published galaxy formation models, 
those of \citet{baugh05} and \citet{bower06}. The models were 
originally tuned to reproduce a subset of observations of the local 
galaxy population and also enjoy notable successes at high redshift. 
We presented the first comparison of the model predictions for the 
properties of \ha\ emitters, extending the work of \citet{dell1,dell2}
and \citet{orsi08} who looked at the nature of 
Lyman-alpha emitters in the models. Observations of \ha\ emitters 
are still in their infancy and the datasets are small. The model 
predicitions bracket the current observational estimates of the 
luminosity function of emitters and are in reasonable agreement with 
the distribution of equivalent widths.  

The next step towards making predicitions of the effectiveness of 
future redshift surveys is to construct mock catalogues from the 
galaxy formation models \citep[see ][]{baugh08}. 
Using the currently available data, we used  
various approaches to fine tune the models to reproduce the observations 
as closely as possible. 
The main technique was to rescale the line and continuum luminosities 
of model galaxies; 
another approach was to randomly dilute or sample galaxies from 
the catalogue. This allowed us to better match the number of observed 
galaxies. The resulting mocks gave reasonable matches to the available 
clustering data around $z \sim 2$. Our goal in this paper was to make 
faithful mock catalogues. The nature of \ha\ emitters in hierarchical 
models will be pursued in a future paper. 

The ability of future surveys to measure the large scale structure of
the Universe can be quantified in terms of their effective volumes. 
The effective volume takes into account the effect of the discreteness 
of sources on the measurment of galaxy clustering. If the discreteness 
noise is comparable to the clustering signal, it becomes hard to extract 
any useful clustering information. Once this point is reached, although 
the available geometrical volume is increased by going deeper in redshift, 
in practice there is little point as no further statistical power is being 
added to the clustering measurments. The error on a power spectrum or 
correlation function measurement scales as the inverse square root 
of the effective volume. In the case of flux-limited samples, the 
number density of sources falls rapidly with increasing redshift beyond 
the median redshift. 
Even though the effective bias of these galaxies tends to increase with 
redshift, it does not do so at a rate sufficient to offset the decline 
in the number density. The \galform\ model naturally predicts the abundance 
and clustering strength of galaxies needed to compute the effective volume 
of a galaxy survey. 

The differences in the expected performance
of \ha\ and H-band selected galaxies when measuring the power 
spectrum is related to the different nature of the galaxies 
selected by these two methods. \ha\ emitters are active star forming
galaxies, which makes them have smaller bias compared to 
H-band selected galaxies. Their redshift distribution is also
very sensitive to the details of the physics of star formation:
The effect of a top-heavy IMF in bursts in the \bau\ model boosts 
the number density of bright emitters, making the redshift 
distrubtion of \ha\ emitters very flat and slowly decreasing towards
high redshifts, in contrast to the predictions of the \bow\ model, 
where a sharp peak at $z\sim 0.5$ and a rapid decrease for higher redshifts
is found. H-band galaxies are less sensitive to this effect, and 
the redshift distributions are similar in both models. This is why the balance between the 
power spectrum amplitude (given by the effective bias) and the
number density is translated in two different effective volumes for
\ha\ and H-band selected galaxies.

Although there are differences in detail between the model 
predictions, they give similar bottom lines for the effective 
volumes of the survey configurations of each galaxy selection.
Comparing the spectroscopic solutions in Table~\ref{table.veff}, a slit 
based survey down to $H_{\rm AB}=22$ would sample 4-10 times the effective volume which 
could be reached by a slitless survey to $\logfha = -15.4$, taking 
into account the likely redshift success rate. To match the performance 
of the H-band survey, an \ha\ survey would need to go much deeper in 
flux, down to $\logfha = -16$. 

We have also looked at the accuracy with which \ha\ emitters and 
H-band selected galaxies will be able to measure the bulk motions 
of galaxies and hence the rate at which fluctuations are growing, 
another key test of gravity and the nature of dark energy. All of the 
samples we considered showed a small systematic difference between 
the measured growth rate and the theoretical expectation, at about 
the $1 \sigma$ level. The error on the growth rate from an \ha\ 
survey with $\logfha > -15.4$ was found to be about three times 
larger than that for a sample with $H_{AB}<22$. 

\section*{Acknowledgements}
We thank the \euclid\ NIS team for providing comments on an earlier 
draft of this paper, and for sharing the results of their simulations of 
the sensitivity of the proposed Euclid spectrographs. We are grateful 
to Michele Cirasuolo, Olivier LeFevre and Henry McCracken for allowing 
us to use their observational estimate of the redshift distribution of 
H-band galaxies. We acknowledge helpful discussions on the topic of this 
paper which arose in a series of telecons chaired by Simon Lilly. 
AO gratefully acknowledges a STFC Gemini studentship. CGL acknowledges 
support from the STFC rolling grant for extragalactic astronomy and 
cosmology at Durham. AC and GZ acknowledge the support from the
Agenzia Spaziale Italiana (ASI, contract N. I/058/08/0).We acknowledge
the effort of Andrew Benson, Richard Bower, Shaun Cole, Carlos Frenk and John Helly
in developing the \galform\ code.

\end{document}